\documentclass[journal]{IEEEtran}

\usepackage{amsthm}
\usepackage{float}
\usepackage{cite}										
\usepackage{amsmath}									
\usepackage{graphicx}
\usepackage{epstopdf}
\usepackage{float}
\usepackage{bm,upgreek}
\usepackage{amsfonts}
\usepackage{algorithm,algpseudocode}
\usepackage{enumitem}
\usepackage{mathtools}
\usepackage{multirow}
\usepackage{booktabs}
\usepackage[subnum]{cases}
\usepackage{setspace}
\usepackage{color}
\usepackage[caption=false,labelformat=simple]{subfig}

\usepackage{pgfplots}
\pgfplotsset{compat=1.5}
\usepackage{pgfplotstable}
\usepgfplotslibrary{statistics}
\pgfplotsset{grid style={dotted,gray}}
\usepackage{tikz}

\allowdisplaybreaks
\emergencystretch=\maxdimen
\ifCLASSINFOpdf
\else
\fi
\pgfplotsset{legend image with text/.style={legend image code/.code={%
\node[anchor=west, align=right] at (0.0cm,0cm) {#1};}},}

\makeatletter
\def\myline{\pgfutil@ifnextchar[{\my@line}{\my@line[]}}%
\def\my@line[#1](#2)(#3){%
\tikz[overlay] \draw[#1]  (#2)--(#3); 
}%
\algrenewcommand\algorithmicindent{1.0em}%

\newtheorem{example}{Example}

\makeatletter
\renewcommand{\ALG@beginalgorithmic}{\small}
\makeatother



\usepackage{etoolbox}
\BeforeBeginEnvironment{algorithmic}{\kern 0.3\baselineskip}
\AfterEndEnvironment{algorithmic}{\kern -0.1\baselineskip}

\usepackage[keeplastbox]{flushend}
\newtheorem{theorem}{Theorem}
\newtheorem{remark}{Remark}

\makeatletter
\renewcommand{\ALG@beginalgorithmic}{\footnotesize}
\makeatother

\pgfplotsset{
    box plot/.style={
        /pgfplots/.cd,
        black,
        only marks,
        mark=-,
        mark size=\pgfkeysvalueof{/pgfplots/box plot width},
        /pgfplots/error bars/y dir=plus,
        /pgfplots/error bars/y explicit,
        /pgfplots/table/x index=\pgfkeysvalueof{/pgfplots/box plot x index},
    },
    box plot box/.style={
        /pgfplots/error bars/draw error bar/.code 2 args={%
            \draw  ##1 -- ++(\pgfkeysvalueof{/pgfplots/box plot width},0pt) |- ##2 -- ++(-\pgfkeysvalueof{/pgfplots/box plot width},0pt) |- ##1 -- cycle;
        },
        /pgfplots/table/.cd,
        y index=\pgfkeysvalueof{/pgfplots/box plot box top index},
        y error expr={
            \thisrowno{\pgfkeysvalueof{/pgfplots/box plot box bottom index}}
            - \thisrowno{\pgfkeysvalueof{/pgfplots/box plot box top index}}
        },
        /pgfplots/box plot
    },
    box plot top whisker/.style={
        /pgfplots/error bars/draw error bar/.code 2 args={%
            \pgfkeysgetvalue{/pgfplots/error bars/error mark}%
            {\pgfplotserrorbarsmark}%
            \pgfkeysgetvalue{/pgfplots/error bars/error mark options}%
            {\pgfplotserrorbarsmarkopts}%
            \path ##1 -- ##2;
        },
        /pgfplots/table/.cd,
        y index=\pgfkeysvalueof{/pgfplots/box plot whisker top index},
        y error expr={
            \thisrowno{\pgfkeysvalueof{/pgfplots/box plot box top index}}
            - \thisrowno{\pgfkeysvalueof{/pgfplots/box plot whisker top index}}
        },
        /pgfplots/box plot
    },
    box plot bottom whisker/.style={
        /pgfplots/error bars/draw error bar/.code 2 args={%
            \pgfkeysgetvalue{/pgfplots/error bars/error mark}%
            {\pgfplotserrorbarsmark}%
            \pgfkeysgetvalue{/pgfplots/error bars/error mark options}%
            {\pgfplotserrorbarsmarkopts}%
            \path ##1 -- ##2;
        },
        /pgfplots/table/.cd,
        y index=\pgfkeysvalueof{/pgfplots/box plot whisker bottom index},
        y error expr={
            \thisrowno{\pgfkeysvalueof{/pgfplots/box plot box bottom index}}
            - \thisrowno{\pgfkeysvalueof{/pgfplots/box plot whisker bottom index}}
        },
        /pgfplots/box plot
    },
    box plot median/.style={
        /pgfplots/box plot,
        /pgfplots/table/y index=\pgfkeysvalueof{/pgfplots/box plot median index},
        semithick,black
    },
    box plot width/.initial=1em,
    box plot x index/.initial=0,
    box plot median index/.initial=1,
    box plot box top index/.initial=2,
    box plot box bottom index/.initial=3,
    box plot whisker top index/.initial=4,
    box plot whisker bottom index/.initial=5,
}

\newcommand{\boxplot}[2][]{
    \addplot [box plot median,#1] table {#2};
    \addplot [forget plot, box plot box,#1] table {#2};
    \addplot [forget plot, box plot top whisker,#1] table {#2};
    \addplot [forget plot, box plot bottom whisker,#1] table {#2};
}

\makeatletter
\IEEEtriggercmd{\fontsize{7.15pt}{8.40pt}\selectfont}
\makeatother
\IEEEtriggeratref{1}

\begin{document}

\newlist{myitemize}{itemize}{3}
\setlist[myitemize,1]{label=\textbullet,leftmargin=6.5mm}

\title{Distributed Gauss-Newton Method for State Estimation Using Belief Propagation}

\author{Mirsad~Cosovic,~\IEEEmembership{Student Member,~IEEE,}
        Dejan Vukobratovic,~\IEEEmembership{Senior Member,~IEEE}

\thanks{M. Cosovic is with Schneider Electric DMS NS, Novi Sad, Serbia (e-mail: mirsad.cosovic@schneider-electric-dms.com). D. Vukobratovic is with Department of Power, Electronic and Communications Engineering, University of Novi Sad, Novi Sad, Serbia (e-mail: dejanv@uns.ac.rs). Demo source code available online at https://github.com/mcosovic.}}

\markboth{}%
{Shell \MakeLowercase{\textit{et al.}}: Bare Demo of IEEEtran.cls for IEEE Journals}

\maketitle

\begin{abstract}

We present a novel distributed Gauss-Newton method for the non-linear state estimation (SE) model based on a probabilistic inference method called belief propagation (BP). The main novelty of our work comes from applying BP sequentially over a sequence of linear approximations of the SE model, akin to what is done by the Gauss-Newton method. The resulting iterative Gauss-Newton belief propagation (GN-BP) algorithm can be interpreted as a distributed Gauss-Newton method with the same accuracy as the centralized SE, however, introducing a number of advantages of the BP framework. The paper provides extensive numerical study of the GN-BP algorithm, provides details on its convergence behavior, and gives a number of useful insights for its implementation.
\end{abstract}


\IEEEpeerreviewmaketitle

\section{Introduction}

\textbf{Motivation:} Electric power systems consist of generation, transmission and consumption spread over wide geographical areas. They are operated from control centers by the power system operators. Maintaining normal operation conditions is of the central importance for the power system operators \cite{abur}. Control centers are traditionally operated in centralized and independent fashion. However, increase in the system size and complexity, as well as external socio-economic factors, lead to deregulation of power systems, resulting in decentralized structure with distributed control centers. Cooperation in control and monitoring across distributed control centers is critical for efficient system operation. Consequently, existing centralized algorithms have to be redefined based on new requirements for distributed operation, scalability and computational efficiency \cite{wu}.

System monitoring is an essential part of control centers, providing control and optimization features that rely on accurate state estimation (SE). The centralized SE approach applies centralized SE algorithms over the measurements collected at the control center. Typically, the Gauss-Newton method is applied to solve the non-linear weighted least-squares (WLS) problem \cite{monticelli}. In contrast, decentralized SE applies distributed SE algorithms in order to distribute communication and computation across multiple control centers. Distributed SE algorithms may or may not require local control centers to coordinate and exchange data with a global control center \cite{gupta}. Their main target is achieving the same state estimate accuracy as the centralized SE algorithms, with as low communication, storage and computation complexity. 

\textbf{Literature Review:} The mainstream distributed SE algorithms exploit matrix decomposition techniques applied over the Gauss-Newton method. These algorithms usually achieve the same accuracy as the centralized SE algorithm and work either with global control center \cite{korres, jiang, aburali} or without it \cite{minot, marelli, tai, reza}. Recently, SE algorithms based on distributed optimization \cite{conejo}, and in particular, the alternating direction method of multipliers became very popular\cite{giannakis, kekatos}. In \cite{anna}, the robust decentralized Gauss-Newton algorithm is proposed which provides flexible communication model, but suffers from slight performance degradation compared to the centralized SE. The work  in \cite{poor} presents a fully distributed SE algorithm for wide-area monitoring which provably converges to the centralized SE. Recently, in \cite{guo}, a new hierarchical multi-area SE method is proposed, where the algorithm converges close to the centralized SE solution with improved convergence speed. We refer the reader to \cite{gomeztax} for a detailed survey of the distributed multi-area SE. In addition, we note that most of the distributed SE papers implicitly consider wide-area monitoring and transmission grid scenario, which is the approach we follow in this paper.

\textbf{Belief-Propagation Approach}: In this paper, we solve the SE problem using probabilistic graphical models, a powerful tool for modeling the dependencies among the systems of random variables. We represent the SE problem using graphical models called factor graphs and solve it using the belief propagation (BP) algorithm. BP is a fully distributed algorithm suitable for accommodation of distributed power sources and time-varying loads. Moreover, placing the SE into the graphical models framework enables efficient inference, but also, a rich collection of tools for learning parameters of the graphical model from observed data \cite{bishop}.

The work in \cite{kavcic} provides the first demonstration of BP applied to the SE problem. Although this work is elaborate in terms of using, e.g., environmental correlation via historical data, it applies BP to a linear approximation of the non-linear functions. The non-linear model is recently addressed in \cite{ilic}, where tree-reweighted BP is applied using preprocessed weights obtained by randomly sampling the space of spanning trees. The work in \cite{fu} investigates Gaussian BP convergence for the DC model. Although the above results provide initial insights on using BP for distributed SE, the BP-based solution for non-linear SE model and the corresponding performance and convergence analysis is still missing. This paper intends to fill this gap.

\textbf{Contributions:} In this paper, we present a novel distributed BP-based Gauss-Newton algorithm, where the BP is applied sequentially over the non-linear model, akin to what is done by the Gauss-Newton method. The resulting Gauss-Newton BP (GN-BP) algorithm represents a BP counterpart of the Gauss-Newton method and introduces a number of advantages over the current state-of-the-art in non-linear SE:
\begin{itemize}[leftmargin=*]
\item The GN-BP is the first BP-based solution for the non-linear SE model achieving exactly the same accuracy as the centralized SE via Gauss-Newton method.
\item In comparison with the distributed SE algorithms that exploit matrix decomposition, the GN-BP is robust to ill-conditioned scenarios caused by significant differences between measurement variances, thus allowing inclusion of arbitrary number of pseudo-measurements without impact to the solution within the observable islands.
\item Due to the sparsity of the underlying factor graph, the GN-BP algorithm has optimal computational complexity (linear per iteration), making it particularly suitable for solving large-scale systems.
\item The GN-BP can be easily designed to provide \emph{asynchronous} operation and integrated as part of the \emph{real-time} systems where newly arriving measurements are processed as soon as they are received \cite{fastDC}. 
\item The GN-BP can easily integrate new measurements: the arrival of a measurement at the control center will define a new factor node which will be seamlessly integrated in the graph as part of the time continuous process. 
\item In the multi-area scenario, the GN-BP algorithm can be implemented over the non-overlapping multi-area SE scenario without the central coordinator, where the GN-BP algorithm neither requires exchanging measurements nor local network topology among the neighboring areas.
\item The GN-BP algorithm is flexible and easy to \emph{distribute} and \emph{parallelize}. Thus, even if implemented in the framework of centralized SE, it can be flexibly matched to distributed computation resources (e.g., parallel processing on graphical-processing units).
\end{itemize}
Finally, we note that this paper significantly extends the conference version \cite{ac_pre}, providing a novel and detailed convergence analysis of the GN-BP algorithm, a novel BP-based bad data analysis, and extensive and insightful numerical results section providing useful recipes for practical implementation. 

\section{SE in Electric Power Systems}
The SE algorithm estimates values of the state variables based on the knowledge of network topology and parameters, and measurements collected across the power system. The network topology and parameters are provided by the network topology processor in the form of the bus/branch model, with branches of the grid usually described using the two-port $\pi$-model \cite[Ch.~1,2]{abur}. As an input, the SE requires a set of measurements $\mathcal{M}$ of different electrical quantities spread across the power network. Using the bus/branch model and available measurements, the observability analysis defines observable and unobservable parts of the network, subsequently defining the additional set of pseudo-measurements needed to determine the solution. Finally, the measurement model can be described as the system of equations \cite[Ch.~4]{abur}:
		\begin{equation}
        \begin{aligned}
        \mathbf{z}=\mathbf{h}(\mathbf{x})+\mathbf{u},
        \end{aligned}
		\label{SE_model}
		\end{equation}
where $\mathbf {x}=[x_1,\dots,x_{n}]^{\mathrm{T}}$ is the vector of the state variables, $\mathbf{h}(\mathbf{x})=$ $[h_1(\mathbf{x})$, $\dots$, $h_k(\mathbf{x})]^{\mathrm{T}}$ is the vector of measurement functions,  $\mathbf{z} = [z_1,\dots,z_k]^{\mathrm{T}}$ is the vector of measurement values, and $\mathbf{u} = [u_1,\dots,u_k]^{\mathrm{T}}$ is the vector of uncorrelated measurement errors. The SE problem in transmission grids is commonly an overdetermined system of equations $(k>n)$ \cite{monticelli}. In general, the system \eqref{SE_model} contains the set of non-linear equations. In a usual scenario, the SE model takes bus voltage magnitudes and bus voltage angles, transformer magnitudes of turns ratio and transformer angles of turns ratio as state variables $\mathbf x$. Without loss of generality, in the rest of the paper, we observe bus voltage angles $\bm \uptheta=$ $[\theta_1,\dots,\theta_N]$ and bus voltage magnitudes $\mathbf V=$ $[V_1,\dots, V_N]$ as state variables $\mathbf x \equiv[\bm \uptheta,\mathbf V]^{\mathrm{T}}$, thus the number of state variables is $n=2N$.

Each measurement $M_i \in \mathcal{M}$ is associated with measured value $z_i$, measurement error  $u_i$, and measurement function $h_i(\mathbf{x})$. Assuming that measurement errors $u_i$ follow a zero-mean Gaussian distribution, the probability density function associated with the \textit{i}-th measurement equals:
		\begin{equation}
        \begin{gathered}
        \mathcal{N}(z_i|\mathbf{x},v_i) = \cfrac{1}{\sqrt{2\pi v_i}} 
        \exp\Bigg\{\cfrac{[z_i-h_i(\mathbf{x})]^2}{2v_i}\Bigg\},
        \end{gathered}
		\label{SE_Gauss_mth}
		\end{equation}
where $v_i$ is the variance of the measurement error $u_i$, and the measurement function $h_i(\mathbf{x})$ connects the vector of state variables $\mathbf{x}$ to the value of the \textit{i}-th measurement.

The solution of the SE problem can be found via maximization of the likelihood function $\mathcal{L}(\mathbf{z}|\mathbf{x})$, which is defined via likelihoods of $k$ independent measurements:  
		\begin{equation}
        \begin{gathered}
		\hat{\mathbf x}=
		\mathrm{arg} \max_{\mathbf{x}}\mathcal{L}(\mathbf{z}|\mathbf{x})=
		\mathrm{arg} \max_{\mathbf{x}}  
		\prod_{i=1}^k \mathcal{N}(z_i|\mathbf{x},v_i).
        \end{gathered}
		\label{SE_likelihood}
		\end{equation}

It can be shown that the solution of \eqref{SE_likelihood} can be obtained by solving the WLS optimization problem \cite[Ch.~2]{abur}. Based on the available set of measurements, the WLS estimator $\hat{\mathbf x} \equiv[\hat{\bm \uptheta},\hat{\mathbf V}]^{\mathrm{T}}$, can be found using the Gauss-Newton method:	
		\begin{subequations}
        \begin{gather}  
		\Big[\mathbf J (\mathbf x^{(\nu)})^\mathrm{T} \mathbf W 
		\mathbf J (\mathbf x^{(\nu)})\Big] \Delta \mathbf x^{(\nu)} =
		\mathbf J (\mathbf x^{(\nu)})^\mathrm{T}
		\mathbf W\mathbf r (\mathbf x^{(\nu)})\label{AC_GN_increment}\\
		\mathbf x^{(\nu+1)} = 
		\mathbf x^{(\nu)} + \Delta \mathbf x^{(\nu)}, \label{AC_GN_update}
        \end{gather}
        \label{AC_GN}%
		\end{subequations}
where $\nu = \{0,1,\dots,\nu_{\max}\}$ is the iteration index and $\nu_{\max}$ is the number of iterations, $\Delta \mathbf x^{(\nu)} \in \mathbb {R}^{n}$ is the vector of increments of the state variables, $\mathbf J (\mathbf x^{(\nu)})\in \mathbb {R}^{k\mathrm{x}n}$ is the Jacobian matrix of measurement functions $\mathbf h (\mathbf x^{(\nu)})$ at $\mathbf x=\mathbf x^{(\nu)}$, $\mathbf{W}\in \mathbb {R}^{k\mathrm{x}k}$ is a diagonal matrix containing inverses of measurement variances, and $\mathbf r (\mathbf x^{(\nu)}) =$ $\mathbf{z}$ $-\mathbf h (\mathbf x^{(\nu)})$ is the vector of residuals. Under these assumptions, the maximum likelihood and WLS estimator are equivalent to the maximum a posteriori (MAP) solution \cite[Sec.~8.6]{barber}.

\section{BP-Based Distributed Gauss-Newton Method}
As the main contribution of this paper, we adopt different methodology to derive efficient BP-based SE method. 

\subsection{Gauss-Newton Method as a Sequential MAP Problem}
Consider the Gauss-Newton method \eqref{AC_GN} where, at each iteration step $\nu$, the algorithm returns a new estimate of $\mathbf{x}$ denoted as $\mathbf{x}^{(\nu)}$. Note that, after a given iteration, an estimate $\mathbf{x}^{(\nu)}$ is a vector of known (constant) values. If the Jacobian matrix $\mathbf J (\mathbf x^{(\nu)})$ has a full column rank, the equation \eqref{AC_GN_increment} represents the linear WLS solution of the minimization problem \cite[Ch.~9]{hansen}:   
		\begin{equation}
        \begin{gathered}
		\min_{\Delta \mathbf x^{(\nu)}} 
		||\mathbf W^{1/2}[\mathbf r (\mathbf x^{(\nu)}) - 
		\mathbf J (\mathbf x^{(\nu)})\Delta \mathbf x^{(\nu)}]||_2^2.
        \end{gathered}
		\label{GN_WLS_increment}
		\end{equation}
Hence, at each iteration $\nu$, the Gauss-Newton method produces WLS solution of the following system of linear equations:  
		\begin{equation}
        \begin{aligned}
        \mathbf r (\mathbf x^{(\nu)})=\mathbf{g}(\Delta \mathbf x^{(\nu)})
        +\mathbf{u},
        \end{aligned}
		\label{GN_linear_SE_model}
		\end{equation}	
where $\mathbf{g}(\Delta \mathbf x^{(\nu)})= \mathbf J (\mathbf x^{(\nu)})\Delta \mathbf x^{(\nu)}$ comprises linear functions, while $\mathbf{u}$ is the vector of measurement errors. The equation \eqref{AC_GN_increment} is the weighted normal equation for the minimization problem defined in \eqref{GN_WLS_increment}, or alternatively \eqref{AC_GN_increment} is a WLS solution of \eqref{GN_linear_SE_model}.
Consequently, the probability density function associated with the \textit{i}-th measurement (i.e., the \textit{i}-th residual component $r_i$) at any iteration step $\nu$ is:
		\begin{multline}
		\mathcal{N}(r_i(\mathbf x^{(\nu)})|{\Delta \mathbf x^{(\nu)}},v_i)\\ 
		= 
         \cfrac{1}{\sqrt{2\pi v_i}} 
        \exp\Bigg\{\cfrac{[r_i(\mathbf x^{(\nu)}) - 
        g_i(\Delta \mathbf x^{(\nu)})]^2}{2v_i}\Bigg\}.
		\label{GN_m_th_residual}
		\end{multline}

The MAP solution of \eqref{SE_likelihood} can be redefined as an iterative optimization problem where, instead of solving \eqref{AC_GN}, we solve:
		\begin{subequations}
        \begin{align}
        \Delta \hat {\mathbf x}^{(\nu)}&=
		\mathrm{arg} \max_{\Delta\mathbf{x}^{(\nu)}}
		\mathcal{L}\Big(\mathbf{r}(\mathbf{x}^{(\nu)})|
		\Delta\mathbf{x}^{(\nu)}\Big)
		 \nonumber \\
		&= \mathrm{arg} \max_{\Delta\mathbf{x}^{(\nu)}} 
		\prod_{i=1}^k \mathcal{N} \Big(r_i(\mathbf{x}^{(\nu)})|
		\Delta\mathbf{x}^{(\nu)},v_i\Big)
        \label{GN_sub_MAP}\\
		\mathbf{x}^{{(\nu+1)}} &= \mathbf{x}^{(\nu)}+ \Delta \hat {\mathbf x}^{(\nu)}.
		\label{GN_MAP_update}
        \end{align}
		\label{GN_MAP}%
		\end{subequations}
In the following, we show that the solution of the above problem \eqref{GN_MAP} can be efficiently obtained using the BP algorithm applied over the underlying factor graph. 

The solution $\Delta \hat {\mathbf x}^{(\nu)}$ in each iteration $\nu = \{0,1, \dots, \nu_{\max}\}$ of the outer iteration loop, is obtained by applying the iterative BP algorithm within inner iteration loops. Every inner BP iteration loop $\tau(\nu) = \{0,1, \dots, \tau_{\max}(\nu)\}$ outputs $\Delta \hat {\mathbf x}^{(\nu,\tau_{\max}(\nu))}$ $\equiv$ $\Delta \hat {\mathbf x}^{(\nu)}$, where $\tau_{\max}(\nu)$ is the number of inner BP iterations within the outer iteration $\nu$. Note that, in general, the BP algorithm operating within inner iteration loops represents an instance of a loopy Gaussian BP over a linear model defined by linear functions $\mathbf{g}(\Delta \mathbf x^{(\nu)})$. Thus, if it converges, it provides a solution equal to the linear WLS solution $\Delta {\mathbf x}^{(\nu)}$ of \eqref{AC_GN_increment}. 

\subsection{The Factor Graph Construction}
From the factorization of the likelihood expression \eqref{GN_sub_MAP}, one easily obtains the factor graph corresponding to the GN-BP method as follows. The increments $\Delta \mathbf x$ of state variables $\mathbf x$ determine the set of variable nodes $\mathcal{V} = \{(\Delta \theta_1, \Delta V_1), \dots, (\Delta \theta_N, \Delta V_N)\}$ and each likelihood function $\mathcal{N} (r_i(\mathbf{x}^{(\nu)})|\Delta\mathbf{x}^{(\nu)},v_i)$ represents the local function associated with the factor node. Since the residual equals $r_i(\mathbf{x}^{(\nu)}) = z_i - h_i(\mathbf{x}^{(\nu)})$, in general, the set of factor nodes $\mathcal{F} =\{f_1,\dots,f_k\}$ is defined by the set of measurements $\mathcal{M}$. The factor node $f_i$ connects to the variable node $\Delta x_s \in \{\Delta \theta_s,\Delta V_s \}$ if and only if the increment of the state variable $\Delta x_s$ is an argument of the corresponding function ${g_i}({\Delta \mathbf x})$, i.e., if the state variable $ x_s \in \{ \theta_s, V_s \}$ is an argument of the measurement function $h_i(\mathbf x)$.

\subsection{Derivation of BP Messages}
\begin{figure}[ht]
	\centering
	\includegraphics[width=4.3cm]{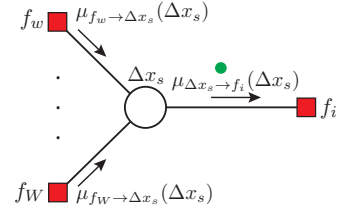}
	\caption{Message $\mu_{\Delta x_s \to f_i}(\Delta x_s)$ from variable node 
	$\Delta x_s$ to factor node $f_i$.}
	\label{Fig_v_f}
	\end{figure}

\textbf{Message from a Variable Node to a Factor Node:} Consider a part of a factor graph shown in Fig. \ref{Fig_v_f} with a group of factor nodes $\mathcal{F}_s=\{f_i,f_w,...,f_W\}$ $\subseteq$ $\mathcal{F}$ that are neighbours of the variable node $\Delta x_s$ $\in$ $\mathcal{V}$. Let us assume that the incoming messages $\mu_{f_w \to \Delta x_s}( \Delta x_s)$, $\dots$, $\mu_{f_W \to \Delta x_s}(\Delta x_s)$ into the variable node $\Delta x_s$ are Gaussian and represented by their mean-variance pairs $(r_{f_w \to \Delta x_s},v_{f_w \to \Delta x_s})$, $\dots$, $(r_{f_W \to \Delta x_s},v_{f_W \to \Delta x_s})$. 
 
 \noindent
The message $\mu_{\Delta x_s \to f_i}(\Delta x_s)$ from the variable node $\Delta x_s$ to the factor node $f_i$ is equal to the product of all incoming factor node to variable node messages arriving at all the other incident edges \cite[Sec.~8.4.4]{bishop}. It is easy to show that the message $\mu_{\Delta x_s \to f_i}(\Delta x_s)$ is proportional to: 
		\begin{equation}
        \begin{gathered}
        \mu_{\Delta x_s \to f_i}(\Delta x_s) \propto
        \mathcal{N}(\Delta x_s|r_{\Delta x_s \to f_i}, v_{\Delta x_s \to f_i}),
        \end{gathered}
		\label{GN_Gauss_vf}
		\end{equation}	
with mean $r_{\Delta x_s \to f_i}$ and variance $v_{\Delta x_s \to f_i}$ obtained as:
		\begin{subequations}
        \begin{align}
         r_{\Delta x_s \to f_i} &= 
        \Bigg( \sum_{f_a \in \mathcal{F}_s\setminus f_i} \cfrac{r_{f_{a} \to \Delta x_s}}
        {v_{f_{a} \to \Delta x_s}}\Bigg)
        v_{\Delta x_s \to f_i}
        \label{GN_vf_mean}\\
		 \cfrac{1}{v_{\Delta x_s \to f_{i}}} &= 
		\sum_{f_a \in \mathcal{F}_s\setminus f_{i}} \cfrac{1}{v_{f_{a} \to \Delta x_s}},
		\label{GN_vf_var}
        \end{align}
		\label{GN_vf_mean_var}%
		\end{subequations}
where $\mathcal{F}_s \setminus f_i$ represents the set of factor nodes incident to the variable node $\Delta x_s$, excluding the factor node $f_i$. To conclude, after the variable node $\Delta x_s$ receives the messages from all of the neighbouring factor nodes from the set $\mathcal{F}_s\setminus f_i$, it evaluates the message $\mu_{\Delta x_s \to f_i}(\Delta x_s)$ and sends it to the factor node $f_i$. 

\begin{figure}[ht]
	\centering
	\includegraphics[width=4.6cm]{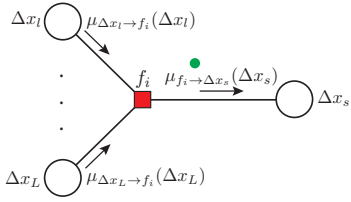}
	\caption{Message $\mu_{f_i \to \Delta x_s}(\Delta x_s)$ from factor 
	node $f_i$ to variable node $\Delta x_s$.}
	\label{Fig_f_v}
	\end{figure}

\textbf{Message from a Factor Node to a Variable Node:} Consider a part of a factor graph shown in Fig. \ref{Fig_f_v} that consists of a group of variable nodes $\mathcal{V}_i =$ $\{\Delta x_s,$ $\Delta x_l,$ $...,$ $\Delta x_L\}$ $\subseteq$ $\mathcal V$ that are neighbours of the factor node $f_i$ $\in$ $\mathcal{F}$. Let us assume that the messages $\mu_{\Delta x_l \to f_i}(\Delta x_l)$, $\dots$, $\mu_{\Delta x_L \to f_i}(\Delta x_L)$  into factor nodes are Gaussian, represented by their mean-variance pairs $(r_{\Delta x_l \to f_i}, v_{\Delta x_l \to f_i})$, $\dots$, $(r_{\Delta x_L \to f_i}, v_{\Delta x_L \to f_i})$. 
	 \noindent

The Gaussian function associated to the factor node $f_i$ is:
		\begin{multline}
		\mathcal{N}(r_i|\Delta x_s,\Delta x_l,\dots,
		\Delta x_L, v_i)\\ 
		\propto 
        \exp\Bigg\{\cfrac{[r_i-g_i
        (\Delta x_s,\Delta x_l,\dots,\Delta x_L)]^2}
        {2v_i}\Bigg\},
		\label{BP_Gauss_measurement_fun}
		\end{multline}
where the model contains only linear functions which we represent in a general form as:
		\begin{equation}
        \begin{gathered}
		g_i(\cdot) =
		C_{\Delta x_s} \Delta x_s + 
		\sum_{\Delta x_b \in \mathcal{V}_i\setminus \Delta x_s} 
		C_{\Delta x_b} \Delta x_b,
		\end{gathered}
		\label{BP_general_measurment_fun}
		\end{equation}
where $\mathcal{V}_i\setminus \Delta x_s$ is the set of variable nodes incident to the factor node $f_i$, excluding the variable node $\Delta x_s$. 

The message $\mu_{f_i \to \Delta x_s}(\Delta x_s)$ from the factor node $f_i$ to the variable node $\Delta x_s$ is defined as a product of all incoming variable node to factor node messages arriving at other incident edges, multiplied by the function associated to the factor node $f_i$, and marginalized over all of the variables associated with the incoming messages \cite[Sec.~8.4.4]{bishop}. It can be shown that the message $\mu_{f_i \to \Delta x_s}(\Delta x_s)$ from the factor node $f_i$ to the variable node $\Delta x_s$ is represented by the Gaussian function:
		\begin{equation}
        \begin{gathered}
        \mu_{f_i \to \Delta x_s}(\Delta x_s) \propto
        \mathcal{N}(\Delta x_s|r_{f_i \to \Delta x_s}, v_{f_i \to \Delta x_s}),
        \end{gathered}
		\label{GN_Gauss_fv}
		\end{equation}
with mean $r_{f_i \to \Delta x_s}$ and variance $v_{f_i \to \Delta x_s}$ obtained as:	
		\begin{subequations}
		\begingroup\makeatletter\def\f@size{9}\check@mathfonts			
		\begin{align}
		r_{f_i \to \Delta x_s}&=
		\cfrac{1}{C_{\Delta x_s}}
		\Bigg( r_i - \sum_{\Delta x_b \in \mathcal{V}_i \setminus \Delta x_s} 
		C_{\Delta x_b} \cdot r_{\Delta x_b \to f_i}  
		 \Bigg)
        \label{GN_fv_mean}	\\
		v_{f_i \to \Delta x_s} &= 
		\cfrac{1}{C_{\Delta x_s}^2}
		\Bigg( v_i + \sum_{\Delta x_b \in \mathcal{V}_i \setminus \Delta x_s} 
		C_{\Delta x_b}^2 \cdot v_{\Delta x_b \to f_i} 
		 \Bigg).
		\label{GN_fv_var}		
        \end{align}
        \endgroup
        \label{GN_fv_mean_var}%
		\end{subequations}	
The coefficients $C_{\Delta x_p},\; \Delta x_p \in \mathcal{V}_i$, are Jacobian elements of the measurement function associated with the factor node $f_i$: 
		\begin{equation}
        \begin{gathered}
		C_{\Delta x_p}=\cfrac{\partial h_i(x_s,x_l,\dots, x_L)}{\partial x_p}.	
        \end{gathered}
		\label{GN_fv_coeff}
		\end{equation}	
			
To summarize, after the factor node $f_i$ receives the messages from all of the neighbouring variable nodes from the set $\mathcal{V}_i\setminus \Delta x_s$, it evaluates the message $\mu_{f_i \to \Delta x_s}(\Delta x_s)$, and sends it to the variable node $\Delta x_s$.  
		
\begin{figure}[ht]
	\centering
	\includegraphics[width=4.3cm]{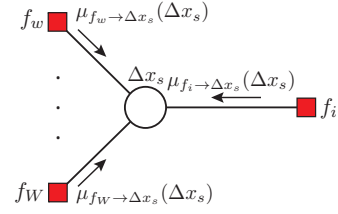}
	\caption{Marginal inference of the variable node $\Delta x_s$.}
	\label{Fig_marginal}
	\end{figure}

\textbf{Marginal Inference:} The marginal of the variable node $\Delta x_s$, illustrated in Fig. \ref{Fig_marginal}, is obtained as the product of all incoming messages into the variable node $\Delta x_s$ \cite[Sec.~8.4.4]{bishop}. 
	 \noindent
It can be shown that the marginal of the state variable $\Delta x_s$ is represented by the Gaussian function: 
\begin{equation}
        \begin{gathered}
        p(\Delta x_s) \propto 
        \mathcal{N}(\Delta x_s|\Delta \hat x_s,v_{\Delta x_s}), 
        \end{gathered}
		\label{GN_Gauss_marg}
		\end{equation}
with mean $\Delta \hat x_s$ which represents the estimated value of the state variable increment $\Delta x_s$ and variance $v_{\Delta x_s}$:
		\begin{subequations}
        \begin{align}
        \Delta \hat x_s &= 
        \Bigg( \sum_{f_c \in \mathcal{F}_s} \cfrac{r_{f_{c} \to \Delta x_s}}
        {v_{f_{c} \to \Delta x_s}}\Bigg)
        v_{\Delta x_s}
        \label{GN_marginal_mean} \\
        \cfrac{1}{v_{\Delta x_s}} &= 
		\sum_{f_c \in \mathcal{F}_s} \cfrac{1}{v_{f_{c} \to \Delta x_s}}, 
		\label{GN_marginal_var}        
        \end{align}
        \label{GN_marginal_mean_var}%
		\end{subequations}
where $\mathcal{F}_s$ is the set of factor nodes incident to the variable node $\Delta x_s$.		

Note that due to the fact that variable node and factor node processing preserves ``Gaussianity" of the messages, each message exchanged in BP is completely represented using only two values: the mean and the variance \cite{ping}.

\subsection{Iterative GN-BP Algorithm}
To present the algorithm precisely, we introduce different types of factor nodes. The \emph{indirect factor nodes} $\mathcal{F}_{\mathrm{ind}} \subset \mathcal{F}$ correspond to measurements that measure state variables indirectly (e.g., power flows and injections). The \emph{direct factor nodes} $\mathcal{F}_{\mathrm{dir}} \subset \mathcal{F}$ correspond to the measurements that measure state variables directly (e.g., voltage magnitudes). Besides direct and indirect factor nodes, we define two additional types of singly-connected factor nodes. The \emph{slack factor node} corresponds to the slack or reference bus where the voltage angle has a given value, therefore, the residual of the corresponding state variable is equal to zero, and its variance tends to zero. Finally, the \emph{virtual factor node} is a singly-connected factor node used if the variable node is not directly measured. Residuals of virtual factor nodes approach zero, while their variances tend to infinity. 

\begin{algorithm} [t]
\caption{The GN-BP}
\label{GN}
\begin{spacing}{1.25}
\begin{algorithmic}[1]
\Procedure {Initialization $\nu=0$}{}
  \For{Each $x_s \in \mathcal{X}$}
  	\State initialize value of $x_s^{(0)}$
  \EndFor 
\EndProcedure
\myline[black](-1.7,1.4)(-1.7,0.3)
\Procedure {Outer iteration loop $\nu=0,1,2,\dots$; $\tau = 0$}{}
\While{stopping criterion for the outer loop is not met} 
\For{Each $f_s \in \mathcal{F}_{\mathrm{dir}}$}
  	\State compute $r_{s}^{(\nu)} = z_s - x_s^{(\nu)}$
  	  \EndFor 
  	  \For{Each $f_s \in \mathcal{F}_{\mathrm{loc}}$}
  	\State send $\mu_{f_s \to \Delta x_s}^{(\nu)}$, $x_s^{(\nu)}$ 
  	to incident $\Delta x_s \in \mathcal{V}$
  \EndFor 
    
  \For{Each $\Delta x_s \in \mathcal{V}$}
  	\State send $\mu_{\Delta x_s \to f_i}^{(\nu){(\tau = 0)}} = \mu_{f_s \to \Delta x_s}^{(\nu)}$,  $x_s^{(\nu)}$ to incident $f_i \in \mathcal{F}_{\mathrm{ind}}$
  \EndFor
  \For{Each $f_i \in \mathcal{F}_{\mathrm{ind}}$}
  \State compute $r_{i}^{(\nu)} = z_i - h_i(\mathbf{x}^{(\nu)})$ and $C_{i,\Delta x_p}^{(\nu)}$; $\Delta x_p \in \mathcal{V}_i$
  \EndFor

\Procedure {Inner Iteration loop $\tau=1,2,\dots$}{} 
\While{stopping criterion for the inner loop is not met} 
  \For{Each $f_i \in \mathcal{F}_{\mathrm{ind}}$}
 	\State compute $\mu_{f_i \to \Delta x_s}^{(\tau)}$ using \eqref{GN_fv_mean_var}
  \EndFor
   \For{Each $\Delta x_s \in \mathcal{V}$}
  \State compute $\mu_{\Delta x_s \to f_i}^{(\tau)}$ using \eqref{GN_vf_mean_var}
  \EndFor
  \EndWhile 
\EndProcedure
\myline[black](-1.7,3.4)(-1.7,0.3)
  \For{Each $\Delta x_s \in \mathcal{V}$}
  	\State compute $\Delta \hat x_s^{(\nu)}$ using \eqref{GN_marginal_mean_var} 		       and $x_s^{(\nu+1)} =  x_s^{(\nu)}+\Delta \hat x_s^{(\nu)}$
  \EndFor
 \EndWhile 
\EndProcedure
\myline[black](-1.7,10.9)(-1.7,0.3)
 \end{algorithmic}
 \end{spacing}
\end{algorithm} \noindent

We refer to direct factor nodes and two additional types of singly-connected factor nodes as local factor nodes $\mathcal{F}_{\mathrm{loc}} \subset \mathcal{F}$. Local factor nodes repeatedly send the same message to incident variable nodes. It is important to note that local factor nodes send messages represented by a triplet: mean (of the residual), variance and the state variable value.

The GN-BP algorithm is presented in Algorithm \ref{GN}, where the set of state variables is defined as $\mathcal{X}=\{x_1,...,x_n\}$. After the initialization (lines 1-5), the outer loop starts by computing residuals for direct and indirect factor nodes, as well as the Jacobian elements, and passes them to the inner iteration loop (lines 8-19). The inner iteration loop (lines 20-29) represents the main algorithm routine which includes BP-based message inference described in the previous subsection. We use synchronous scheduling, where all messages in a given inner iteration are updated  using the output of the previous iteration as an input \cite{elidan}. The output of the inner iteration loop is the estimate of the state variable increments. Finally, the outer loop updates the set of state variables (lines 30-32). The outer loop iterations are repeated until the stopping criteria is met.

\begin{example}[Constructing a factor graph] In this toy example, using a simple 3-bus model presented in Fig. \ref{Fig_ex_bus_branch_5meas}, we demonstrate the conversion from a bus/branch model with a given measurement configuration into the corresponding factor graph. 
	\begin{figure}[ht]
	\centering
	\begin{tabular}{@{}c@{}}
	\subfloat[]{\label{Fig_ex_bus_branch_5meas}
	\includegraphics[width=2.5cm]{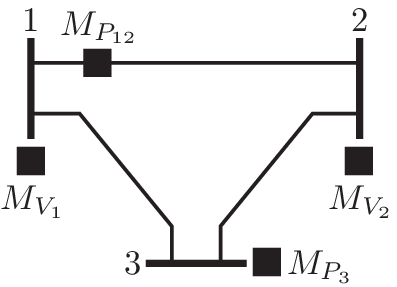}}
	\end{tabular}\quad
	\begin{tabular}{@{}c@{}}
	\subfloat[]{\label{Fig_ex_FG_5meas}
	\includegraphics[width=5.1cm]{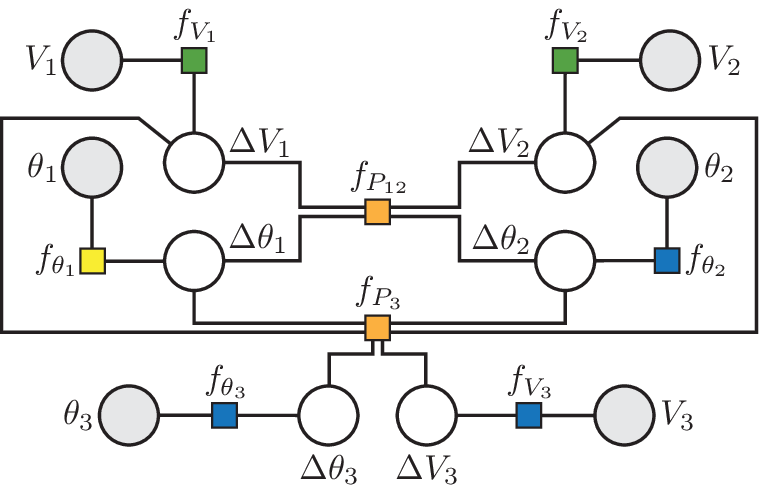}} 
	\end{tabular}
	\caption{Transformation of the bus/branch model and
	 measurement configuration (subfigure a) into the corresponding 
	 factor graph with different types of factor nodes (subfigure b).}
	\label{Fig_ex_diff_factor}
	\end{figure} 
	
The corresponding factor graph is given in Fig. \ref{Fig_ex_FG_5meas}, where the set of state variables is $\mathcal{X} =$ $\{(\theta_1,V_1),$  $(\theta_2,V_2),$ $(\theta_3,V_3)\}$ and the set of variable nodes is $\mathcal{V} =$ $\{(\Delta \theta_1,\Delta V_1),$ $(\Delta \theta_2, \Delta V_2),$ $(\Delta \theta_3,\Delta V_3)\}$. The indirect factor nodes (orange squares) are defined by corresponding measurements, where in our example, active power flow $M_{P_{12}}$ and active power injection $M_{P_3}$ measurements  are mapped into factor nodes $\mathcal{F}_{\mathrm{ind}} = $ $\{f_{P_{12}},$ $f_{P_{3}} \}$. The set of local factor nodes $\mathcal{F}_{\mathrm{loc}}$ consists of the set of direct factor nodes (green squares) $\mathcal{F}_{\mathrm{dir}} = $ $\{f_{V_{1}},$ $f_{V_{2}} \}$ defined by bus voltage magnitude measurements $M_{V_{1}}$ and $M_{V_2}$, virtual factor nodes (blue squares) and the slack factor node (yellow square). 	  
\end{example}

\subsection{Discussion}
The presented GN-BP algorithm can be easily adapted to the multi-area SE model. Therein, each area runs the GN-BP algorithm in a fully parallelized way, exchanging messages asynchronously with neighboring areas. The algorithm may run as a continuous process, with each new measurement being seamlessly processed by the distributed state estimator. The BP approach is robust to ill-conditioned scenarios caused by significant differences between measurement variances, thus alleviating the need for observability analysis. Indeed, one can include arbitrarily large set of additional pseudo-measurements initialized using extremely high variances without affecting the BP solution within the observable part of the system \cite{fastDC}.

\section{Convergence Analysis}
In this part, we present convergence analysis of the GN-BP algorithm with synchronous scheduling, and propose an improved GN-BP algorithm that applies synchronous scheduling \emph{with randomized damping}. We emphasize that the convergence of the GN-BP algorithm critically depends on the convergence behavior of each of the inner iteration loops.  

\subsection{Synchronous Scheduling}

In the following, it will be useful to consider a subgraph of the factor graph that contains the set of variable nodes $\mathcal{V}$, the set of indirect factor nodes $\mathcal{F}_{\mathrm{ind}} = \{f_1, \dots, f_m \} \subset \mathcal{F}$, and the set of edges $\mathcal{B} \subseteq \mathcal{V} \times \mathcal{F}_{\mathrm{ind}}$ connecting them. The number of edges in this subgraph is $b = |\mathcal{B}|$. Within the subgraph, we will consider a factor node $f_i \in \mathcal{F}_{\mathrm{ind}}$ connected to its neighboring set of variable nodes $\mathcal{V}_i = \{\Delta x_q, \dots, \Delta x_Q \} \subset \mathcal{V}$ by a set of edges $\mathcal{B}_i = \{b_i^q, \dots,b_i^Q \}  \subset \mathcal{B}$, where $d_i = |\mathcal{V}_i|$ is the degree of $f_i$. Next, we provide results on convergence of both variances and means of inner iteration loop messages, respectively.
	
\textbf{Convergence of the Variances:} From \eqref{GN_vf_var} and \eqref{GN_fv_var}, we note that the evolution of variances is independent of mean values of messages and measurements. Let $\mathbf {v}_{\mathrm{s}} \in \mathbb{R}^b$ denote a vector of variance values of messages from indirect factor nodes $\mathcal{F}_{\mathrm{ind}}$ to variable nodes $\mathcal{V}$. Substituting \eqref{GN_vf_var} in \eqref{GN_fv_var}, the variance updates take the recursive form $\mathbf {v}_{\mathrm{s}}^{(\tau)} = f \big( \mathbf {v}_{\mathrm{s}}^{(\tau-1)}\big)$. More precisely, using simple matrix algebra, one can obtain the evolution of the variances $\mathbf {v}_{\mathrm{s}}$ in the following matrix form:		
		\begin{equation}
        \begin{aligned}
        \mathbf {v}_{\mathrm{s}}^{(\tau)} = 
        \Big[\big(\mathbf{\widetilde C}^{-1}
        \bm {\Pi} \mathbf{\widetilde C}\big)\cdot
        \big(\mathfrak{D}(\mathbf{A}) 
        \big)^{-1}
        +\bm {\Sigma}_{\mathrm{a}}\mathbf{\widetilde C}^{-1}
        \Big] \mathbf{i}, 
        \end{aligned}
        \label{con_5}        
		\end{equation}
where $\mathbf{\widetilde C} = \mathbf{C}\mathbf{C}^{\mathrm{T}}$ and $\mathbf{A} = \mathbf{\Gamma} \bm {\Sigma}_{\mathrm{s}}^{-1} \mathbf{\Gamma}^\mathrm{T} + \mathbf{L}$. Note that in \eqref{con_5}, the dependence on $\mathbf {v}_{\mathrm{s}}^{(\tau-1)}$ is hidden in matrix $\mathbf{A}$, or more precisely, in matrix $\mathbf{{\Sigma}_{\mathrm{s}}}$. For brevity, we describe vectors, matrices and matrix-operators involved in \eqref{con_5} in Appendix.		
		
\begin{theorem}
The variances $\mathbf {v}_{\mathrm{s}}$ from indirect factor nodes to variable nodes always converge to a unique fixed point $\lim_{\tau \to \infty} \mathbf {v}_{\mathrm{s}}^{(\tau)} =\hat{\mathbf {v}}_{\mathrm{s}}$ for any initial point $\mathbf {v}_{\mathrm{s}}^{(\tau=0)} > 0$.
\end{theorem}
\begin{proof}\renewcommand{\qedsymbol}{}
The theorem can be proved by showing that $f \big( \mathbf {v}_{\mathrm{s}}\big)$ satisfies the conditions of the so-called standard function \cite{hanly}, following similar steps as in the proof of Lemma 1 in \cite{zhang}.
\end{proof}

\textbf{Convergence of the Means:}
Equations \eqref{GN_vf_mean} and \eqref{GN_fv_mean} show that the evolution of the mean values depends on the variance values. Due to Theorem 1, it is possible to simplify evaluation of mean values $\mathbf {r}_{\mathrm{s}}$ from indirect factor nodes $\mathcal{F}_{\mathrm{ind}}$ to variable nodes $\mathcal{V}$ by using the fixed-point values of $\hat{\mathbf {v}}_{\mathrm{s}}$. The evolution of means $\mathbf {r}_{\mathrm{s}}$ becomes a set of linear equations: 
		\begin{equation}
        \begin{aligned}
        \mathbf {r}_{\mathrm{s}}^{(\tau)} = \mathbf{\widetilde r}
        -  \bm \Omega 
        \mathbf {r}_{\mathrm{s}}^{(\tau-1)},
        \end{aligned}
        \label{rand_mean}
		\end{equation}
where $\mathbf{\widetilde r} =  \mathbf{C}^{-1} \mathbf{r}_{\mathrm{a}} - \mathbf{D} \cdot \big(\mathfrak{D} (\hat {\mathbf{A}})\big)^{-1} \cdot \mathbf{L}\mathbf{r}_\mathrm{b}$, $\bm \Omega =\mathbf{D}\cdot\big(\mathfrak{D} (\hat {\mathbf{A}})\big)^{-1} \cdot \mathbf{\Gamma}\hat{\bm \Sigma}_{\mathrm{s}}^{-1}$, $ \hat {\mathbf{A}} = \mathbf{\Gamma} \hat{\bm \Sigma}_{\mathrm{s}}^{-1} \mathbf{\Gamma}^\mathrm{T} + \mathbf{L}$ and $\mathbf{D} = \mathbf{C}^{-1}\mathbf{\Pi} \mathbf{C}$ (as above, we describe vectors, matrices and matrix-operators involved in \eqref{rand_mean} in Appendix).
		
\begin{theorem}
The means $\mathbf {r}_{\mathrm{s}}$ from indirect factor nodes to variable nodes converge to a unique fixed point $\lim_{\tau \to \infty} \mathbf {r}_{\mathrm{s}}^{(\tau)} =\hat {\mathbf {r}}_{\mathrm{s}}:$
		\begin{equation}
        \begin{aligned}
        \hat {\mathbf {r}}_{\mathrm{s}} =\big(\mathbf{I}+\bm \Omega \big)^{-1} 
        \mathbf{\widetilde r},
        \label{fixed}
        \end{aligned}
		\end{equation}
for any initial point $\mathbf {r}_{\mathrm{s}}^{(\tau=0)}$ if and only if the spectral radius $\rho(\bm \Omega)<1$.
\end{theorem}
\begin{proof}\renewcommand{\qedsymbol}{}
The proof follows steps in Theorem 5.2 \cite{hanly}.
\end{proof}

To summarize, the convergence of the inner iteration loop of the GN-BP algorithm depends on the spectral radius of the matrix $\bm \Omega$. If the spectral radius $\rho(\bm \Omega)<1$, the GN-BP algorithm in the inner iteration loop $\nu$ will converge and the resulting vector of mean values will be equal to the solution of the MAP estimator. Consequently, the convergence of the GN-BP with synchronous scheduling in each outer iteration loop $\nu$ depends on the spectral radius of the matrix:
		\begin{multline}
		\bm \Omega(\mathbf x^{(\nu)}) = \big[\mathbf{C}(\mathbf x^{(\nu)})^{-1}
		\mathbf{\Pi} \mathbf{C}(\mathbf x^{(\nu)}) \big]\\
		\cdot\big[\mathfrak{D} (\mathbf{\Gamma} \hat{\bm \Sigma}_{\mathrm{s}}^{-1}
        \mathbf{\Gamma}^\mathrm{T} + \mathbf{L})\big]^{-1} \cdot \big(\mathbf{\Gamma}
        \hat{\bm \Sigma}_{\mathrm{s}}^{-1}\big). 
		\label{GN_syn_omega_con}
		\end{multline}
		
\begin{remark}		
The GN-BP with synchronous scheduling converges to a unique fixed point if and only if $\rho_{\mathrm{syn}}<1$, where:
		\begin{equation}
        \begin{aligned}
  		\rho_{\mathrm{syn}} = 
        \max\{\rho \big(\bm {\Omega}({\mathbf x}^{(\nu)}):
        \nu = 0,1,\dots,\nu_{\max}\}. 
        \end{aligned}
        \label{maxsy}
		\end{equation}
\end{remark}		

\subsection{Synchronous Scheduling with Randomized Damping}
Next, we propose an improved GN-BP algorithm that applies synchronous scheduling with randomized damping. Several previous works reported that damping the BP messages improves the convergence of BP\cite{zhang, pretti}. Here, we propose a different randomized damping approach, where each mean value message from indirect factor node to a variable node is damped independently with probability $p$, otherwise, the message is calculated as in the standard GN-BP algorithm. The damped message is evaluated as a linear combination of the message from the previous and the current iteration, with weights $\alpha_1$ and $1-\alpha_1$, respectively. 

Using the proposed damping, equation \eqref{rand_mean} is redefined as:
		\begin{equation}
        \begin{aligned}
        \mathbf {r}_{\mathrm{d}}^{(\tau)} = \mathbf {r}_{\mathrm{q}}^{(\tau)}
        +\alpha_1 \mathbf {r}_{\mathrm{r}}^{(\tau-1)} + 
        \alpha_2\mathbf {r}_{\mathrm{r}}^{(\tau)},
        \label{rand_3}
        \end{aligned}
		\end{equation}
where $0<\alpha_1<1$ is the weighting coefficient, and $\alpha_2 = 1 - \alpha_1$. In the above expression, 
$\mathbf {r}_{\mathrm{q}}^{(\tau)}$ and $\mathbf {r}_{\mathrm{r}}^{(\tau)}$ are obtained as: 
		\begin{subequations}
        \begin{align}
        \mathbf {r}_{\mathrm{q}}^{(\tau)} &= 
        \mathbf {Q} \mathbf{\widetilde r}
        - \mathbf {Q}  \bm \Omega 
        \mathbf {r}_{\mathrm{s}}^{(\tau-1)}
        \label{rand_1}\\
         \mathbf {r}_{\mathrm{r}}^{(\tau)} &= 
        \mathbf {R}\mathbf{\widetilde r}
        - \mathbf {R} \bm \Omega
        \mathbf {r}_{\mathrm{s}}^{(\tau-1)},
        \label{rand_2}
        \end{align}
		\end{subequations} 
where diagonal matrices $\mathbf {Q} \in \mathbb{F}_2^{b \times b}$ and $\mathbf {R} \in \mathbb{F}_2^{b \times b}$ are defined as $\mathbf {Q} = \mathrm{diag}(1 - q_1,...,1 - q_b)$, $q_i \sim \mathrm{Ber}(p)$, and $\mathbf {R} = \mathrm{diag}(q_1,...,q_b)$, respectively, and where $\mathrm{Ber}(p) \in \{0,1\}$ is a Bernoulli random variable with probability $p$ independently sampled for each mean value message. 	

Substituting \eqref{rand_1} and \eqref{rand_2} in \eqref{rand_3}, we obtain:
		\begin{equation}
        \begin{aligned}
        \mathbf {r}_{\mathrm{d}}^{(\tau)} = \big(\mathbf {Q}+  \alpha_2 \mathbf {R}\big)
        \mathbf{\widetilde r} - 
        \big(\mathbf {Q} \bm \Omega + \alpha_2\mathbf {R} 
        \bm \Omega - \alpha_1\mathbf {R} \big) 
        \mathbf {r}_{\mathrm{s}}^{(\tau-1)}. 
        \label{rand_4}
        \end{aligned}
		\end{equation}
Note that $\mathbf {r}_{\mathrm{r}}^{(\tau-1)} = \mathbf {R} \mathbf {r}_{\mathrm{s}}^{(\tau-1)}$. In a more compact form \eqref{rand_4} can be written as follows: 
		\begin{equation}
        \begin{aligned}
        \mathbf {r}_{\mathrm{d}}^{(\tau)} =\mathbf{\bar r} - \bm {\bar \Omega} 
        \mathbf {r}_{\mathrm{s}}^{(\tau-1)}, 
        \label{rand_5}
        \end{aligned}
		\end{equation}
where $\mathbf{\bar r} = \big(\mathbf {Q}+  \alpha_2 \mathbf {R}\big)  \mathbf{\widetilde r}$ and $\bm {\bar \Omega} = \mathbf {Q} \bm \Omega + \alpha_2\mathbf {R} \bm \Omega - \alpha_1\mathbf {R}$.
		
\begin{theorem}
The means $\mathbf {r}_{\mathrm{d}}$ from indirect factor nodes to variable nodes converge to a unique fixed point $\hat{\mathbf {r}}_{\mathrm{d}} = \lim_{\tau \to \infty} \mathbf {r}_{\mathrm{d}}^{(\tau)}$ for any initial point $\mathbf {r}_{\mathrm{d}}^{(\tau=0)}$ if and only if the spectral radius $\rho(\bm {\bar \Omega})<1$. Furthermore, for the resulting fixed point $\hat{\mathbf {r}}_{\mathrm{d}}$, it holds that $\hat{\mathbf {r}}_{\mathrm{d}} = \hat{\mathbf {r}}_{\mathrm{s}}$. 
\end{theorem}	

\begin{proof}\renewcommand{\qedsymbol}{} The proof can be found in the Appendix. \end{proof}

To summarize, the convergence of the GN-BP with randomized damping in every outer iteration loop $\nu$ is governed by the spectral radius of the matrix:
		\begin{equation}
        \begin{gathered}
		\bm {\bar \Omega}(\mathbf x^{(\nu)}) = \mathbf {Q} 
		\bm \Omega(\mathbf x^{(\nu)}) +
		\alpha_2\mathbf {R} 
        \bm \Omega(\mathbf x^{(\nu)}) - \alpha_1\mathbf {R}.	
        \end{gathered}
		\label{GN_rand_con}
		\end{equation}

\begin{remark}			
The GN-BP with randomized damping will converge to a unique fixed point if and only if $\rho_{\mathrm{rd}}<1$, where:
		\begin{equation}
        \begin{aligned}
  		\rho_{\mathrm{rd}} = 
        \max\{\rho \big(\bm {\bar \Omega}({\mathbf x}^{(\nu)}):
        \nu = 0,1,\dots,\nu_{\max}\}, 
        \end{aligned}
        \label{maxrd}
		\end{equation}
and the resulting fixed point is equal to the fixed point obtained by the GN-BP with synchronous scheduling. 
\end{remark}	

In Section VI, we demonstrate that the GN-BP with randomized damping dramatically improves the GN-BP convergence.    

\section{Bad Data Analysis}
Besides the SE algorithm, one of the essential SE routines is the bad data analysis, whose main task is to detect and identify measurement errors, and eliminate them if possible. SE algorithms based on the Gauss-Newton method proceed with the bad data analysis after the estimation process is finished. This is usually done by processing the measurement residuals \cite[Ch.~5]{abur}, and typically, the largest normalized residual test (LNRT) is used to identify bad data \cite{guo}. The LNRT is performed after the Gauss-Newton algorithm converged in the repetitive process of identifying and eliminating bad data measurements one after another \cite{korres}.

Using analogies from the LNRT, we define the bad data test based on the BP messages from factor nodes to variable nodes. The presented model establishes local criteria to detect and identify bad data measurements. In Section VI, we demonstrate that the BP-based bad data test (BP-BDT) significantly improves the bad data detection over the LNRT.

\begin{figure}[ht]
	\centering
	\includegraphics[width=7.0cm]{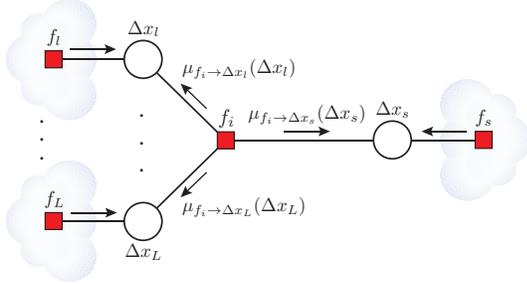}
	\caption{The part of the factor graph with messages 
	from factor node $f_i$ 
	to group of variable nodes $\mathcal{V}_i = \{\Delta x_s, \Delta x_l,..., 
	\Delta x_L\}$.}
	\label{Fig_bad}
	\end{figure}

\textbf{The Belief Propagation Bad Data Test:} Consider a part of the factor graph shown in Fig. \ref{Fig_bad} and focus on a single measurement $M_i \in \mathcal{M}$ that defines the factor node $f_i$ $\in$ $\mathcal{F}$. Factor nodes $\{f_s,$ $f_l,$ $\dots,$ $f_L\}$ carry a collective evidence of the rest of the factor graph about the group of variable nodes $\mathcal{V}_i = \{\Delta x_s, \Delta x_l,..., \Delta x_L\}$ $\subseteq$ $\mathcal V$ incident to $f_i$. 

Assume that the estimation process is done, and the residual of the measurement $M_i$ is given as:  
		\begin{equation}
        \begin{gathered}
        r_i(\mathbf{ x}_i + \Delta {\hat{\mathbf x}}_i) = 
        z_i - h_i(\mathbf{x}_i + \Delta {\hat{\mathbf x}}_i), 
        \end{gathered}
		\label{residual}
		\end{equation}
where $\mathbf{x}_i =$ $[x_s,$ $x_l,$ $\dots,$ $x_L]^{\mathrm{T}}$ is the vector of state variables, while $\Delta {\hat{\mathbf x}}_i =$ $[\Delta \hat{x}_s,$ $\Delta \hat{x}_l,$ $\dots,$ $\Delta \hat{x}_L]^{\mathrm{T}}$ is the corresponding estimate vector of state variable increments. Let us define vectors $\mathbf{r}_{f_i} = $ $[r_{f_i \to \Delta x_s},$ $r_{f_i \to \Delta x_l},$ $\dots,$ $r_{f_i \to \Delta x_L}]^{\mathrm{T}}$ and $\mathbf{v}_{f_i} = $ $[v_{f_i \to \Delta x_s},$ $v_{f_i \to \Delta x_l},$ $\dots,$ $v_{f_i \to \Delta x_L}]^{\mathrm{T}}$ of mean and variance values of BP messages sent from the factor node $f_i$ to the variable nodes in $\mathcal{V}_i$, respectively. 

According to \eqref{GN_marginal_mean}, the vector of state variable increments $\Delta {\hat{\mathbf x}}_i$ is determined as:  
		\begin{equation}
        \begin{gathered}
        \Delta {\hat{\mathbf x}}_i = 
        [\mathrm{diag}(\mathbf{v}_{{\Delta x}_i})]\cdot
        [\mathrm{diag}(\mathbf{v}_{f_i})]^{-1} \cdot \mathbf{r}_{f_i} 
        + \mathbf{b},
        \end{gathered}
		\label{evo}
		\end{equation}
where $\mathbf{v}_{{\Delta x}_i} =$ $[v_{\Delta x_s},$ $v_{\Delta x_l},$ $\dots,$ $v_{\Delta x_L}]^{\mathrm{T}}$ is the vector of variable node variances obtained using \eqref{GN_marginal_var} and the vector $\mathbf{b}$ carries evidence of the rest of the graph about the corresponding variable nodes $\mathcal{V}_i$.

From \eqref{evo}, one can note that the BP-based SE algorithm decomposes the contribution of each factor node to state variable increments, thus providing insight in the structure of measurement residual in \eqref{residual}, where the impact of each measurement can be observed. More precisely, the expression $[\mathrm{diag}(\mathbf{v}_{f_i})]^{-1} \cdot$ $\mathbf{r}_{f_i}$ determines the influence of the measurement $M_i$ to the residual \eqref{residual}. 
To recall, the mean-value messages $\mathbf{r}_{f_i}$ contain ``beliefs'' of the factor node $f_i$ about variable nodes in $\mathcal{V}_i$, with the corresponding variances $\mathbf{v}_{f_i}$. Consequently, if the measurement $M_i$ represents bad data,  it will likely provide an inflated values of the normalized residual components $[\mathrm{diag}(\mathbf{v}_{f_i})]^{-1} \cdot$ $\mathbf{r}_{f_i}$ in \eqref{evo}. Thus, we observe the following vector corresponding to each factor node $f_i$ to detect the bad data:   
		\begin{equation}
        \begin{gathered}
        \mathbf{r}_{\mbox{\scriptsize BP},f_i} = 
        [\mathrm{diag}(\mathbf{v}_{f_i})]^{-1}
        \cdot [\mathrm{diag}(\mathbf{r}_{f_i})] \cdot
        \mathbf{r}_{f_i}. 
        \end{gathered}
		\label{det_bad}
		\end{equation}
Note, the expression $[\mathrm{diag}(\mathbf{r}_{f_i})] \cdot$ $\mathbf{r}_{f_i} = $ $[r_{f_i \to \Delta x_s}^2,$ $r_{f_i \to \Delta x_l}^2,$ $\dots,$ $r_{f_i \to \Delta x_L}^2]^{\mathrm{T}}$ favors larger values of $\mathbf{r}_{f_i}$. 

To summarize, we define the BP-BDT algorithm following similar steps as the LNRT \cite[Sec.~5.7]{abur}. Namely, after the state estimation process is done, we compute $\mathbf{r}_{\mbox{\scriptsize BP},f_i}$, $f_i$ $\in$ $\mathcal{F}$, using \eqref{det_bad}, and observe $\bar {r}_{\mbox{\scriptsize BP},f_i}$ as the largest element of $\mathbf{r}_{\mbox{\scriptsize BP},f_i}$. Comparing $\bar {r}_{\mbox{\scriptsize BP},f_i}$ values among all factor nodes, we find the largest such value ${r}_{\mbox{\scriptsize BP},f_m}$ corresponding to the $m$-th factor node. If ${r}_{\mbox{\scriptsize BP},f_m} > \kappa$, then the $m$-th measurement is suspected as bad data, where $\kappa$ is the bad data identification threshold.

\section{Numerical Results}
\textbf{Simulation Setup:} In the simulated model, we start with a given IEEE test case and apply the power flow analysis to generate the exact solution. Further, we corrupt the exact solution by the additive white Gaussian noise of variance $v$, and we observe the set of measurements: legacy (active and reactive injections and power flows, line current magnitudes and bus voltage magnitudes) and phasor measurement units (bus voltage and line current phasors). The set of measurements is selected in such a way that the system is observable. More precisely, for each scenario, we generate 300 random measurement configurations in order to obtain average performances.

In all models, we use measurement variance equal to $v_i = 10^{-10}\,\mbox{p.u.}$ for PMUs, and $v_i = 10^{-4}\,\mbox{p.u.}$ for legacy devices. To initialize the GN-BP and Gauss-Newton method, we run algorithms using ``flat start" with a small random perturbation \cite[Sec.~9.3]{abur} or ``warm start" where we use the same initial point as the one applied in AC power flow. Finally, randomized damping parameters are set to $p = 0.8$ and $\alpha_1 = 0.4$ (obtained by exhaustive search). To evaluate the performance of the GN-BP algorithm, we convert each of the above randomly generated IEEE test cases with a given measurement configuration into the corresponding factor graph, and we run the GN-BP algorithm.  

\textbf{Convergence and Accuracy:} We consider IEEE 30-bus test case with 5 PMUs and the set of legacy measurements with redundancy $\gamma$ $\in$ $\{2,3,4,5\}$. We first set the number of inner iterations to a high value of $\tau_{\max}(\nu)=5000$ iterations for each outer iteration $\nu$, where $\nu_{\max} = 11$, with the goal of investigating convergence and accuracy of GN-BP.

\begin{figure}[ht]
	\centering
	\captionsetup[subfigure]{oneside,margin={1.3cm,0cm}}
	\begin{tabular}{@{}c@{}}
	\subfloat[]{\label{plot1a}
	\centering
	\begin{tikzpicture}
  	\begin{axis}[width=6.5cm, height=5.0cm,
   	x tick label style={/pgf/number format/.cd,
   	set thousands separator={},fixed},
   	xlabel={Spectral Radius $\rho$},
   	ylabel={Empirical CDF $F(\rho)$},
   	label style={font=\footnotesize},
   	grid=major,
   	legend style={legend pos=north west,font=\scriptsize, column sep=0cm},
	legend columns=2,   	
   	ymin = 0, ymax = 1.1,
   	xmin = 0.55, xmax = 1.25,
   	xtick={0.6,0.7,0.8,0.9,1,1.1,1.2},
   	tick label style={font=\footnotesize},
   	ytick={0,0.1,0.2,0.3,0.4,0.5,0.6,0.7,0.8,0.9,1.0}]
	\addlegendimage{legend image with text=$\rho_{\mathrm{syn}}$}
    \addlegendentry{}
    \addlegendimage{legend image with text=$\rho_{\mathrm{rd}}$}
    \addlegendentry{}   
    
    \addplot[mark=diamond*,mark repeat=70, mark size=1.5pt, blue, dashed] 
   	table [x={x}, y={y}] {./figure/plot1a/01_std10m2red2_sy.txt}; 
   	\addlegendentry{}  
	\addplot[mark=diamond*,mark repeat=70, mark size=1.5pt, blue] 
   	table [x={x}, y={y}] {./figure/plot1a/02_std10m2red2_rd.txt};
   	\addlegendentry{$\gamma = 2$}
   	
   	\addplot[mark=otimes*, mark repeat=68, mark size=1.5pt, red, dashed]
	table [x={x}, y={y}] {./figure/plot1a/03_std10m2red3_sy.txt};
   	\addlegendentry{}
	\addplot[mark=otimes*, mark repeat=68, mark size=1.5pt, red]
	table [x={x}, y={y}] {./figure/plot1a/04_std10m2red3_rd.txt};   	
   	\addlegendentry{$\gamma = 3$}
   	
	\addplot[mark=square*,mark repeat=58, mark size=1.4pt, orange, dashed]
   	table [x={x}, y={y}] {./figure/plot1a/05_std10m2red4_sy.txt};
   	\addlegendentry{}
   	\addplot[mark=square*,mark repeat=58, mark size=1.4pt, orange]
   	table [x={x}, y={y}] {./figure/plot1a/06_std10m2red4_rd.txt};
   	\addlegendentry{$\gamma = 4$}
   	
    \addplot[mark=triangle*,mark repeat=74, mark size=1.5pt, black, dashed]
   	table [x={x}, y={y}] {./figure/plot1a/07_std10m2red5_sy.txt};
   	\addlegendentry{}
   	\addplot[mark=triangle*,mark repeat=74, mark size=1.5pt, black]
   	table [x={x}, y={y}] {./figure/plot1a/08_std10m2red5_rd.txt};
   	\addlegendentry{$\gamma = 5$}
  	\end{axis}
	\end{tikzpicture}}
	\end{tabular}\\
	\begin{tabular}{@{}c@{}}
	\subfloat[]{\label{plot1b}
	\begin{tikzpicture}
	\begin{axis}[width=6.5cm, height=5.0cm,
   	x tick label style={/pgf/number format/.cd,
   	set thousands separator={},fixed},
   	xlabel={Spectral Radius $\rho$},
   	ylabel={Empirical CDF $F(\rho)$},
   	label style={font=\footnotesize},
   	grid=major,
   	legend style={legend pos=north west,font=\scriptsize, column sep=0cm},
	legend columns=2,   	
   	ymin = 0, ymax = 1.1,
   	xmin = 0.55, xmax = 1.25,
   	xtick={0.6,0.7,0.8,0.9,1,1.1,1.2},
   	tick label style={font=\footnotesize},
   	ytick={0,0.1,0.2,0.3,0.4,0.5,0.6,0.7,0.8,0.9,1.0}]
	\addlegendimage{legend image with text=$\rho_{\mathrm{syn}}$}
    \addlegendentry{}
    \addlegendimage{legend image with text=$\rho_{\mathrm{rd}}$}
    \addlegendentry{}   
    
    \addplot[mark=diamond*,mark repeat=70, mark size=1.5pt, blue, dashed] 
   	table [x={x}, y={y}] {./figure/plot1b/01_std10m2red2_sy.txt}; 
   	\addlegendentry{}  
	\addplot[mark=diamond*,mark repeat=70, mark size=1.5pt, blue] 
   	table [x={x}, y={y}] {./figure/plot1b/02_std10m2red2_rd.txt};
   	\addlegendentry{$\gamma = 2$}
   	
   	\addplot[mark=otimes*, mark repeat=70, mark size=1.5pt, red, dashed]
	table [x={x}, y={y}] {./figure/plot1b/03_std10m2red3_sy.txt};
   	\addlegendentry{}
	\addplot[mark=otimes*, mark repeat=70, mark size=1.5pt, red]
	table [x={x}, y={y}] {./figure/plot1b/04_std10m2red3_rd.txt};   	
   	\addlegendentry{$\gamma = 3$}
   	
	\addplot[mark=square*,mark repeat=59, mark size=1.4pt, orange, dashed]
   	table [x={x}, y={y}] {./figure/plot1b/05_std10m2red4_sy.txt};
   	\addlegendentry{}
   	\addplot[mark=square*,mark repeat=59, mark size=1.4pt, orange]
   	table [x={x}, y={y}] {./figure/plot1b/06_std10m2red4_rd.txt};
   	\addlegendentry{$\gamma = 4$}
   	
    \addplot[mark=triangle*,mark repeat=65, mark size=1.5pt, black, dashed]
   	table [x={x}, y={y}] {./figure/plot1b/07_std10m2red5_sy.txt};
   	\addlegendentry{}
   	\addplot[mark=triangle*,mark repeat=65, mark size=1.5pt, black]
   	table [x={x}, y={y}] {./figure/plot1b/08_std10m2red5_rd.txt};
   	\addlegendentry{$\gamma = 5$}   	
  	\end{axis}
	\end{tikzpicture}}
	\end{tabular}
	\caption{The maximum spectral radii $\rho_{\mathrm{syn}}$ with synchronous and $\rho_{\mathrm{rd}}$ with randomized damping 
	scheduling over outer iterations $\nu = \{0,1,2,\dots,12\}$ for legacy redundancy $\gamma$ $\in$ $\{2,3,4,5\}$ and variance $v = 10^{-4}$ for IEEE 30-bus test case using ``flat start" (subfigure a) and ``warm start" (subfigure b).}
	\label{plot1}
	\end{figure}
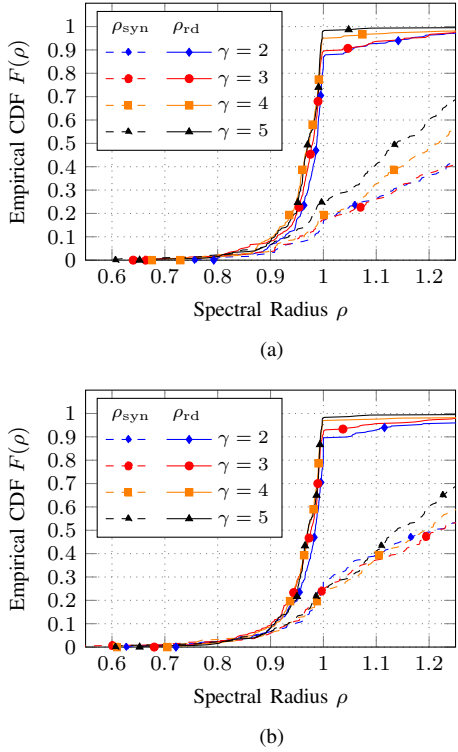

Fig. \ref{plot1} shows empirical cumulative density function (CDF) $F(\rho)$ of spectral radius $\rho_{\mathrm{syn}}$ and $\rho_{\mathrm{rd}}$ for different redundancies for ``flat start" and ``warm start". For each scenario, the randomized damping case is superior in terms of the spectral radius. For example, for redundancy $\gamma = 5$ and ``flat start", we record convergence with probability $0.98$ for randomized damping and $0.25$ for synchronous scheduling. When operated in ``warm start" via, e.g., large-scale historical data, the GN-BP can be integrated into continuous real-time SE framework following similar steps as in \cite{fastDC}.

In the following, we compare the accuracy of the GN-BP algorithm to that of the Gauss-Newton method. We use the weighted residual sum of squares (WRSS) as a metric:
	\begin{equation}
	\begin{gathered}
	\mathrm{WRSS} = \sum_{i=1}^k 
	\cfrac{[z_i-h_i({\mathbf x})]^2}{v_i}.
	\end{gathered}
	\label{num_WRSS}
	\end{equation}
Note that WRSS is the value of the objective function of the optimization problem \cite[Sec.~2.5]{abur} we are solving, thus it is a suitable metric for the SE accuracy. Finally, we normalize the obtained $\mathrm{WRSS}_{\mbox{\scriptsize BP}}^{(\nu)}$ over outer iterations $\nu$ by $\mathrm{WRSS}_{\mbox{\scriptsize WLS}}$ of the centralized SE obtained using the Gauss-Newton method after 12 iterations (which we adopt as a normalization constant). 
	\begin{figure}[ht]
	\centering
	\captionsetup[subfigure]{oneside,margin={1.8cm,0cm}}
	\begin{tabular}{@{}c@{}}
	\subfloat[]{\label{plot2a}
	\centering	
	\begin{tikzpicture}
  	\begin{axis}[width=4cm, height=4.5cm,
   	x tick label style={/pgf/number format/.cd,
   	set thousands separator={},fixed},
   	y tick label style={/pgf/number format/.cd,fixed,
   	fixed zerofill, precision=3, /tikz/.cd},
   	xlabel={Outer iterations  $\nu$},
   	ylabel={$\mathrm{WRSS}_{\mbox{\scriptsize BP}}^{(\nu)}/
   	\mathrm{WRSS}_{\mbox{\scriptsize WLS}}$},
   	label style={font=\footnotesize},
   	grid=major,
   	xtick={1,2,3,4,5},
   	xticklabels={$4$, $5$, $6$, $7$, $8$},
   	tick label style={font=\footnotesize}]
	\addplot [color=blue, only marks, mark=o, line width=0.6pt]
 	plot [error bars, y dir = both, y explicit, 
 	error bar style={line width=0.6pt}]
 	table[x =x, y =y, y error =e]{./figure/plot2/errorbar_red4.txt};
  	\end{axis}
	\end{tikzpicture}}
	\end{tabular}
	\begin{tabular}{@{}c@{}}
	\subfloat[]{\label{plot2b}
	\begin{tikzpicture}
  	\begin{axis}[width=4cm, height=4.5cm,
   	x tick label style={/pgf/number format/.cd,
   	set thousands separator={},fixed},
   	y tick label style={/pgf/number format/.cd,fixed,
   	fixed zerofill, precision=3, /tikz/.cd},
   	xlabel={Outer iterations  $\nu$},
   	ylabel={$\mathrm{WRSS}_{\mbox{\scriptsize BP}}^{(\nu)}/
   	\mathrm{WRSS}_{\mbox{\scriptsize WLS}}$},
   	label style={font=\footnotesize},
   	grid=major,
   	ytick={1,1.001,0.999},
   	xtick={1,2,3,4,5},
   	xticklabels={$4$, $5$, $6$, $7$, $8$},
   	tick label style={font=\footnotesize}]
	\addplot [color=blue, only marks, mark=o, line width=0.6pt]
 	plot [error bars, y dir = both, y explicit, 
 	error bar style={line width=0.6pt}]
 	table[x =x, y =y, y error =e]{./figure/plot2/errorbar_red5.txt};
  	\end{axis}
	\end{tikzpicture}}
	\end{tabular}
	\caption{The GN-BP normalized WRSS (i.e., 
	$\mathrm{WRSS}_{\mbox{\tiny BP}}^{(\nu)}/
   	\mathrm{WRSS}_{\mbox{\tiny WLS}}$) for IEEE 30-bus test case 
   	using ``flat start" and legacy redundancy $\gamma = 4$ (subfigure a) and
   	$\gamma = 5$ (subfigure b).}
	\label{plot2}
	\end{figure}
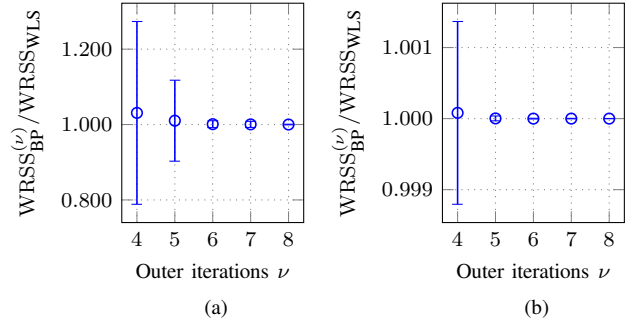 

Fig. \ref{plot2} shows error bar (mean and standard deviation) of normalized WRSS for ``flat start" scenario where redundancy set to $\gamma = 4$ and $\gamma = 5$ within converged simulations. As shown, $(\mathrm{WRSS}_{\mbox{\scriptsize BP}}$ $/$ $\mathrm{WRSS}_{\mbox{\scriptsize WLS}})$ $\to$ $1$, which corresponds to the case where the GN-BP converges to the exactly same solution as the centralized Gauss-Newton method. 

\textbf{Scalability and Complexity:}  Next, we use the mean absolute difference (MAD) between the state variables in two consecutive iterations as a metric:
	\begin{equation}
	\begin{gathered}
	\mathrm{MAD} =  \cfrac{1}{n} \sum_{i=1}^n 
	|\Delta x_i|.
	\end{gathered}
	\label{num_MAD}
	\end{equation}	
The MAD value represents average component-wise shift of the state estimate over the iterations, thus it may be used to quantify the rate of convergence. 

To investigate the rate of convergence as the size of the system increases, we provide MAD values for IEEE 118-bus and 300-bus test case using the ``warm start" and legacy redundancy $\gamma = 4$ with $20$ and $50$ PMUs, respectively. In the following, in order to reduce the number of inner iterations, we define an alternative inner iteration scheme. Namely, as before, we are running algorithm up to $\tau_{\max}(\nu)$, but here we allow interruption of the inner iteration loops when accuracy-based criterion is met. More precisely, the algorithm in the inner iteration loop is running until the following criterion is reached:    
		\begin{equation}
        \begin{gathered}      
		|\mathbf{r}_{f \to \Delta x}^{(\nu,\tau)}-
		\mathbf{r}_{f \to \Delta x}^{(\nu, \tau-1})| < \epsilon(\nu)
		\;\;\mathrm{or}\;\;
		\tau(\nu) = \tau_{\max}(\nu),
		\end{gathered}
		\label{num_break2}
		\end{equation}
where $\mathbf{r}_{f \to \Delta x}$ represents the vector of mean-value messages from factor nodes to variable nodes, $\epsilon(\nu) = [10^{-2},$ $10^{-4},$ $10^{-6},$ $10^{-8},$ $10^{-10}]$ is the threshold at iteration $\nu$. The upper limit on inner iterations is $\tau_{\max}(\nu)=6000$ for each outer iteration $\nu$, where $\nu_{\max} = 4$.

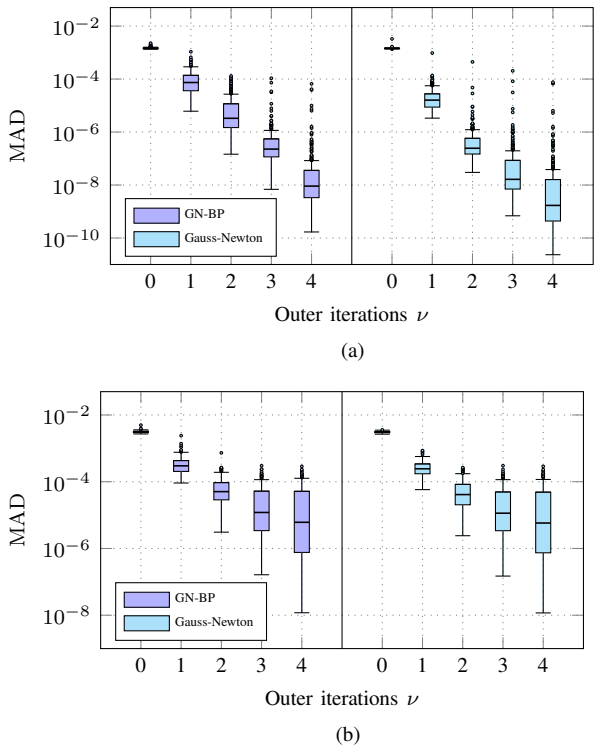
\begin{figure}[ht]
	\captionsetup[subfigure]{oneside,margin={1.6cm,0cm}}
	\begin{tabular}{@{}c@{}}
	\subfloat[]{\label{plot3a}
	\centering
	\begin{tikzpicture}
	\begin{semilogyaxis} [box plot width=1.0mm,
	xlabel={Outer iterations  $\nu$},
   	ylabel={$\mathrm{MAD}$},
   	grid=major,   		
   	xmin=0, xmax=12, ymin=0.00000000001, ymax =0.05,	
   	xtick={1,2,3,4,5,7,8,9,10,11},
   	xticklabels={0, 1, 2, 3, 4, 0, 1, 2, 3, 4},
   	ytick={0, 0.01, 0.0001, 0.000001, 0.00000001, 0.0000000001},
   	width=8cm,height=5cm,
   	tick label style={font=\footnotesize}, label style={font=\footnotesize},
   	legend style={draw=black,fill=white,legend cell align=left,font=\tiny,
   	legend pos=south west}]
	\boxplot [
    forget plot, fill=blue!30,
    box plot whisker bottom index=1,
    box plot whisker top index=5,
    box plot box bottom index=2,
    box plot box top index=4,
    box plot median index=3] {./figure/plot3/ieee118_mad_bp_data.txt};   
    \draw[solid, fill=blue!30, draw=black]
    (axis cs:0.685, 3) rectangle (axis cs:0.965, 4);
	\addlegendimage{area legend,fill=blue!30,draw=black}
	\addlegendentry{GN-BP};
	   
	\boxplot [
    forget plot, fill=cyan!30, 
    box plot whisker bottom index=1,
    box plot whisker top index=5,
    box plot box bottom index=2,
    box plot box top index=4,
    box plot median index=3] {./figure/plot3/ieee118_mad_wls_data.txt};        
	\draw[solid, fill=red!30, draw=black] 
	(axis cs:0.685, 3) rectangle (axis cs:0.965, 4);
	\addlegendimage{area legend,fill=cyan!30,draw=black}
	\addlegendentry{Gauss-Newton}; 
	
    \addplot[only marks, mark options={draw=black, fill=blue!30},mark size=0.5pt] 
	table[x index=0, y index=1] {./figure/plot3/ieee118_mad_bp_outliers.txt};	
	\addplot[only marks, mark options={draw=black, fill=cyan!30},mark size=0.5pt] 
	table[x index=0, y index=1] {./figure/plot3/ieee118_mad_wls_outliers.txt};   
    
	\draw [thin] (60,\pgfkeysvalueof{/pgfplots/ymin}) -- 
	(60,\pgfkeysvalueof{/pgfplots/ymax});

	\end{semilogyaxis}
	\end{tikzpicture}}
	\end{tabular}\\
	\begin{tabular}{@{}c@{}}
	\subfloat[]{\label{plot3b}
	\begin{tikzpicture}
	\begin{semilogyaxis} [box plot width=1.0mm,
	xlabel={Outer iterations  $\nu$},
   	ylabel={$\mathrm{MAD}$},
   	grid=major,   		
   	xmin=0, xmax=12, ymin=0.000000001, ymax =0.05,	
   	xtick={1,2,3,4,5,7,8,9,10,11},
   	xticklabels={0, 1, 2, 3, 4, 0, 1, 2, 3, 4},
   	ytick={0, 0.01, 0.0001, 0.000001, 0.00000001, 0.0000000001},
   	width=8cm,height=5cm,
   	tick label style={font=\footnotesize}, label style={font=\footnotesize},
   	legend style={draw=black,fill=white,legend cell align=left,font=\tiny,
   	legend pos=south west}]
	\boxplot [
    forget plot, fill=blue!30,
    box plot whisker bottom index=1,
    box plot whisker top index=5,
    box plot box bottom index=2,
    box plot box top index=4,
    box plot median index=3] {./figure/plot3/ieee300_mad_bp_data.txt};   
    \draw[solid, fill=blue!30, draw=black]
    (axis cs:0.685, 3) rectangle (axis cs:0.965, 4);
	\addlegendimage{area legend,fill=blue!30,draw=black}
	\addlegendentry{GN-BP};
	   
	\boxplot [
    forget plot, fill=cyan!30, 
    box plot whisker bottom index=1,
    box plot whisker top index=5,
    box plot box bottom index=2,
    box plot box top index=4,
    box plot median index=3] {./figure/plot3/ieee300_mad_wls_data.txt};        
	\draw[solid, fill=cyan!30, draw=black] 
	(axis cs:0.685, 3) rectangle (axis cs:0.965, 4);
	\addlegendimage{area legend,fill=cyan!30,draw=black}
	\addlegendentry{Gauss-Newton}; 
	
    \addplot[only marks, mark options={draw=black, fill=blue!30},mark size=0.5pt] 
	table[x index=0, y index=1] {./figure/plot3/ieee300_mad_bp_outliers.txt};	
	\addplot[only marks, mark options={draw=black, fill=cyan!30},mark size=0.5pt] 
	table[x index=0, y index=1] {./figure/plot3/ieee300_mad_wls_outliers.txt};   
    
	\draw [thin] (60,\pgfkeysvalueof{/pgfplots/ymin}) -- 
	(60,\pgfkeysvalueof{/pgfplots/ymax});

	\end{semilogyaxis}
	\end{tikzpicture}}
	\end{tabular}
	\caption{The MAD values of the GN-BP algorithm and Gauss-Newton method 
	for IEEE 118-bus (subfigure a) and IEEE 300-bus (subfigure b) 
	test case.}
	\label{plot3}
\end{figure}

Fig. \ref{plot3} compares the MAD values of the GN-BP and Gauss-Newton method for IEEE 118-bus and 300-bus test cases within converged simulations. The GN-BP has achieved the presented performance at $\tau_{\max}(\nu) =$ $\{131,$ $488,$ $855,$ $1357,$ $2587\}$ and $\tau_{\max}(\nu) =$ $\{242,$ $1394,$ $5987,$ $6000,$ $6000\}$ (i.e., median values) for IEEE 118-bus and 300-bus test case, respectively. Note that the GN-BP exhibits very similar convergence performance to that of the centralized SE. Note also that it is difficult to directly compare the two, due to a large difference in computational loads of a single (outer) iteration. For example, the complexity of a single iteration remains constant but significant (due to matrix inversion) over iterations for the centralized SE algorithm, while it gradually increases for the GN-BP starting from an extremely low complexity at initial outer iterations. Namely, the overall complexity of the centralized SE scales as $O(n^3)$, and this can be reduced to $O(n^{2+c})$ by employing matrix inversion techniques that exploit the sparsity of involved matrices. The complexity of BP depends on the sparsity of the underlying factor graph, as the computational effort per iteration is proportional to the number of edges in the factor graph. For each of the $k$ measurements, the degree of the corresponding factor node is limited by a (typically small) constant. Indeed, for any type of measurements, the corresponding measurement function depends only on a few state variables corresponding to the buses in the local neighbourhood of the bus/branch where the measurement is taken. As $n$ and $k$ grow large, the number of edges in the factor graph scales as $O(n)$, thus the computational complexity of GN-BP scales linearly \emph{per iteration}. The scaling of the number of BP iterations as $n$ grows large is a more challenging problem. We leave the detailed analysis on the scaling of the number of inner GN-BP iterations per outer iteration for our future work.

\textbf{Bad Data Analysis:} To investigate the proposed BP-BDT, we use IEEE 14-bus and 30-bus test case, with 3 PMUs and 5 PMUs, respectively, and the set of legacy measurements of redundancy $\gamma = 3$. In each of 300 random measurement configurations, we randomly generate a bad measurement among legacy measurements, with variance set to $v_{\mathrm{b}20} = 400v_i$ or $v_{\mathrm{b}40} = 1600v_i$ (i.e., $20\sigma_i$ or $40\sigma_i$). For each simulation, we record only the largest elements ${r}_{\mbox{\scriptsize BP},f_m}$ and $r_{\mbox{\scriptsize N},m}$ obtained using BP-BDT and LNRT, respectively.   

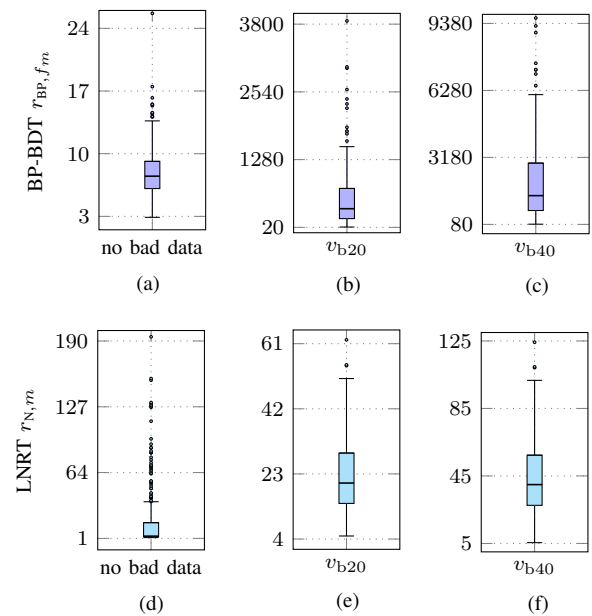
\begin{figure}[ht]
	\captionsetup[subfigure]{oneside,margin={1.0cm,0cm}}
	\begin{tabular}{@{}c@{}}
	\subfloat[]{\label{plot4a}
	\centering
	\begin{tikzpicture}
	\begin{axis} [box plot width=1.0mm,
	xlabel={},
   	ylabel={BP-BDT ${r}_{\mbox{\tiny BP},f_m}$},
   	grid=major,   		
    xmin=0.5, xmax=1.5, ymin=1.5, ymax =26,	
    xtick={1},
   	xticklabels={no bad data},
   	ytick={3, 10, 17, 24},
   	width=3cm,height=4.5cm,
   	tick label style={font=\footnotesize}, label style={font=\footnotesize},
   	legend style={draw=black,fill=white,legend cell align=left,font=\tiny,
   	legend pos=south west}]
	\boxplot [
    forget plot, fill=blue!30,
    box plot whisker bottom index=1,
    box plot whisker top index=5,
    box plot box bottom index=2,
    box plot box top index=4,
    box plot median index=3] {./figure/plot4/4a_dat_ieee14_no_bad_BP.txt};  
    \addplot[only marks, mark options={draw=black, fill=blue!30},
    mark size=0.5pt] 
	table[x index=0, y index=1] {./figure/plot4/4a_out_ieee14_no_bad_BP.txt}; 
	\end{axis}
	\end{tikzpicture}}
	\end{tabular}
	\begin{tabular}{@{}c@{}}
	\subfloat[]{\label{plot4b}
	\begin{tikzpicture}
	\begin{axis} [box plot width=1.0mm,
    /pgf/number format/.cd, use comma, 1000 sep={},
	xlabel={},
   	ylabel={},
   	grid=major,   		
    xmin=0.5, xmax=1.5, ymin=-80, ymax =4000,	
    xtick={1},
   	xticklabels={$v_{\mathrm{b}20}$},
	ytick={20, 1280, 2540, 3800},
   	width=3cm,height=4.5cm,
   	tick label style={font=\footnotesize}, label style={font=\footnotesize},
   	legend style={draw=black,fill=white,legend cell align=left,font=\tiny,
   	legend pos=south west}]
	\boxplot [
    forget plot, fill=blue!30,
    box plot whisker bottom index=1,
    box plot whisker top index=5,
    box plot box bottom index=2,
    box plot box top index=4,
    box plot median index=3] {./figure/plot4/4b_dat_ieee14_1leg_20_BP.txt}; 
    \addplot[only marks, mark options={draw=black, fill=blue!30},
    mark size=0.5pt] 
	table[x index=0, y index=1] {./figure/plot4/4b_out_ieee14_1leg_20_BP.txt};       
	\end{axis}
	\end{tikzpicture}}
	\end{tabular}
	\begin{tabular}{@{}c@{}}
	\subfloat[]{\label{plot4c}
	\begin{tikzpicture}
	\begin{axis} [box plot width=1.0mm,
    /pgf/number format/.cd, use comma, 1000 sep={},
	xlabel={},
   	ylabel={},
   	grid=major,   		
    xmin=0.5, xmax=1.5, ymin=-350, ymax =9800,	
    xtick={1},
   	xticklabels={$v_{\mathrm{b}40}$},
	ytick={80, 3180, 6280, 9380},
   	width=3cm,height=4.5cm,
   	tick label style={font=\footnotesize}, label style={font=\footnotesize},
   	legend style={draw=black,fill=white,legend cell align=left,font=\tiny,
   	legend pos=south west}]
	\boxplot [
    forget plot, fill=blue!30,
    box plot whisker bottom index=1,
    box plot whisker top index=5,
    box plot box bottom index=2,
    box plot box top index=4,
    box plot median index=3] {./figure/plot4/4c_dat_ieee14_1leg_40_BP.txt};
   \addplot[only marks, mark options={draw=black, fill=blue!30},
    mark size=0.5pt] 
	table[x index=0, y index=1] {./figure/plot4/4c_out_ieee14_1leg_40_BP.txt};    
	\end{axis}
	\end{tikzpicture}}
	\end{tabular}\\
	\captionsetup[subfigure]{oneside,margin={1.25cm,0cm}}
	\begin{tabular}{@{\hspace{-0.15cm}}c@{}}
	\subfloat[]{\label{plot4d}
	\centering
	\begin{tikzpicture}
	\begin{axis} [box plot width=1.0mm,
	xlabel={},
   	ylabel={LNRT ${r}_{\mbox{\tiny N},m}$},
   	grid=major,   		
    xmin=0.5, xmax=1.5, ymin=-10, ymax =200,	
    xtick={1},
   	xticklabels={no bad data},
   	ytick={1,64, 127,190},
   	width=3cm,height=4.5cm,
   	tick label style={font=\footnotesize}, label style={font=\footnotesize},
   	legend style={draw=black,fill=white,legend cell align=left,font=\tiny,
   	legend pos=south west}]
	\boxplot [
    forget plot, fill=cyan!30,
    box plot whisker bottom index=1,
    box plot whisker top index=5,
    box plot box bottom index=2,
    box plot box top index=4,
    box plot median index=3] {./figure/plot4/4d_dat_ieee14_no_bad_WLS.txt};   
    \addplot[only marks, mark options={draw=black, fill=cyan!30},
    mark size=0.5pt] 
	table[x index=0, y index=1] {./figure/plot4/4d_out_ieee14_no_bad_WLS.txt}; 
	\end{axis}		
	\end{tikzpicture}}
	\end{tabular}
	\captionsetup[subfigure]{oneside,margin={0.7cm,0cm}}	
	\begin{tabular}{@{\hspace{0.31cm}}c@{}}
	\subfloat[]{\label{plot4e}
	\begin{tikzpicture}
	\begin{axis} [box plot width=1.0mm,
    /pgf/number format/.cd, use comma, 1000 sep={},
	xlabel={},
   	ylabel={},
   	grid=major,   		
    xmin=0.5, xmax=1.5, ymin=1, ymax =65,	
    xtick={1},
   	xticklabels={$v_{\mathrm{b}20}$},
	ytick={4,23,42,61},
   	width=3cm,height=4.5cm,
   	tick label style={font=\footnotesize}, label style={font=\footnotesize},
   	legend style={draw=black,fill=white,legend cell align=left,font=\tiny,
   	legend pos=south west}]
	\boxplot [
    forget plot, fill=cyan!30,
    box plot whisker bottom index=1,
    box plot whisker top index=5,
    box plot box bottom index=2,
    box plot box top index=4,
    box plot median index=3] {./figure/plot4/4e_dat_ieee14_1leg_20_WLS.txt};
    \addplot[only marks, mark options={draw=black, fill=cyan!30},
    mark size=0.5pt] 
	table[x index=0, y index=1] {./figure/plot4/4e_out_ieee14_1leg_20_WLS.txt}; 
	\end{axis}       
	\end{tikzpicture}}
	\end{tabular}
	\captionsetup[subfigure]{oneside,margin={0.9cm,0cm}}		
	\begin{tabular}{@{\hspace{0.14cm}}c@{}}
	\subfloat[]{\label{plot4f}
	\begin{tikzpicture}
	\begin{axis} [box plot width=1.0mm,
    /pgf/number format/.cd, use comma, 1000 sep={},
	xlabel={},
   	ylabel={},
   	grid=major,   		
    xmin=0.5, xmax=1.5, ymin=0, ymax =130,	
    xtick={1},
   	xticklabels={$v_{\mathrm{b}40}$},
	ytick={5,45,85,	125},
   	width=3cm,height=4.5cm,
   	tick label style={font=\footnotesize}, label style={font=\footnotesize},
   	legend style={draw=black,fill=white,legend cell align=left,font=\tiny,
   	legend pos=south west}]
	\boxplot [
    forget plot, fill=cyan!30,
    box plot whisker bottom index=1,
    box plot whisker top index=5,
    box plot box bottom index=2,
    box plot box top index=4,
    box plot median index=3] {./figure/plot4/4f_dat_ieee14_1leg_40_WLS.txt};
    \addplot[only marks, mark options={draw=black, fill=cyan!30},
    mark size=0.5pt] 
	table[x index=0, y index=1] {./figure/plot4/4f_out_ieee14_1leg_40_WLS.txt};        
	\end{axis}
	\end{tikzpicture}}
	\end{tabular}
	\caption{Comparisons between BP-BDT and LNRT for bad data 
	free measurement set (subfigure a and d), a single bad data 
	in the measurement set with
	variance $v_{\mathrm{b}20}$ (subfigure b and e) 
	and $v_{\mathrm{b}40}$ (subfigure c and f)
	for IEEE 14-bus test case using ``warm start".}
	\label{plot4}
\end{figure}

Fig. \ref{plot4} compares the BP-BDT to the LNRT for IEEE 14-bus test case using \emph{``warm start"}. The BP-BDT successfully identified the bad measurement in 291 and 294 cases, while LNRT succeeded in 220 and 240 cases, for $v_{\mathrm{b}20}$ and $v_{\mathrm{b}40}$, respectively. Figs. \ref{plot4b}, \ref{plot4c}, \ref{plot4e} and \ref{plot4f} show observed distributions of BP-BDT and LNRT metrics (${r}_{\mbox{\scriptsize BP},f_m}$ and $r_{\mbox{\scriptsize N},m}$) when tests succeeded in identifying the bad measurement. Clearly, the metric resolution between the cases without bad data (Figs. \ref{plot4a} and \ref{plot4d}) and the cases when the bad data exists in the measurement set, allows easier identification of bad data with the BP-BDT, providing for easier adjustment of the bad data identification threshold $\kappa$, in contrast to the LNRT. 
  
The BP-BDT reconfirmed the improved bad data detection for the case where two bad measurements exist in the measurement set (both with variance $v_{\mathrm{b}20}$ or $v_{\mathrm{b}40}$) for IEEE 30-bus test case initialized via \emph{``flat start"}. The BP-BDT successfully identified one of the two bad data samples after the first cycle (i.e., in the presence of another bad measurement) in 267 and 275 cases, while the LNRT identified the first bad data sample in 222 and 251 cases.

\section{Conclusions}   
In this paper, we presented a novel GN-BP algorithm, which is an efficient and accurate BP-based implementation of the iterative Gauss-Newton method. GN-BP can be highly parallelized and flexibly distributed in the context of multi-area SE. In our ongoing work, we are investigating GN-BP in asynchronous, dynamic and real-time SE with online bad data detection, supported by future 5G communication infrastructure \cite{CMag}. 

\section*{Appendix}
\textbf{Definitions of Vectors, Matrices and Operators Related with Section IV:}
The vector $\mathbf {v}_{\mathrm{s}} \in \mathbb{R}^b$ can be decomposed as $\mathbf {v}_{\mathrm{s}}^{(\tau)} =$ $[\mathbf {v}_{\mathrm{s},1}^{(\tau)},$  $\dots,$ $\mathbf {v}_{\mathrm{s},m}^{(\tau)}]^{\mathrm{T}}$, where the $i$-th element $\mathbf {v}_{\mathrm{s},i} \in \mathbb{R}^{d_i}$ is equal to: $\mathbf {v}_{\mathrm{s},i}^{(\tau)} =$ $[ v_{f_{i} \to \Delta x_q}^{(\tau)},$ $\dots,$ $v_{f_{i} \to \Delta x_Q}^{(\tau)}]$. 

The operator $\mathfrak{D}(\mathbf{A}) \equiv \mathrm{diag}(A_{11}, \dots, A_{bb})$, where $A_{ii}$ is the $i$-th diagonal entry of the matrix $\mathbf{A}$. The all-one vector $\mathbf{i}$ is of dimension $b$ and is equal to $\mathbf{i} = [1,\dots,1]^\mathrm{T}$. The diagonal matrix $\bm {\Sigma}_{\mathrm{s}}$ is obtained as $\bm {\Sigma}_{\mathrm{s}} = \mathrm{diag}\big(\mathbf {v}_{\mathrm{s}}^{(\tau-1)}\big) \in \mathbb{R}^{b \times b}$.

The matrix $\mathbf{C} = \mathrm{diag}\big(\mathbf{C}_{1},\dots,\mathbf{C}_{m}\big) \in \mathbb{R}^{b \times b}$ contains diagonal entries of the Jacobian non-zero elements, where the $i$-th element $\mathbf {C}_i = [C_{\Delta x_q}, \dots, C_{\Delta x_Q}] \in \mathbb{R}^{d_i}$. The matrix $\bm {\Sigma}_{\mathrm{a}} = \mathrm{diag}\big(\bm {\Sigma}_{\mathrm{a,1}},\dots,\bm {\Sigma}_{\mathrm{a},m} \big) \in \mathbb{R}^{b \times b}$ contains indirect factor node variances, with the $i$-th entry $\bm {\Sigma}_{\mathrm{a},i} = [v_i, \dots, v_i] \in \mathbb{R}^{d_i}$. 

The matrix $\mathbf{L} = \mathrm{diag}\big(\mathbf{L}_{1}, \dots, \mathbf{L}_{m} \big) \in \mathbb{R}^{b \times b}$ contains inverse variances from singly-connected factor nodes to a variable node, if such nodes exist, where the $i$-th element $\mathbf{L}_i = \big[l_{x_q}, \dots, {l}_{x_Q} \big] \in \mathbb{R}^{d_i}$, for example, $l_{x_q} =$ $1/{v_{f_{\mathrm{d},q} \to \Delta x_q}}$.

The matrix $\mathbf{\Pi} = \mathrm{diag}\big(\mathbf{\Pi}_1, \dots, \mathbf{\Pi}_m \big) \in \mathbb{F}_2^{b \times b}$, $\mathbb{F}_2 = \{0,1\}$, is a block-diagonal matrix in which the $i$-th element is a block matrix $\mathbf{\Pi}_i = \mathbf{1}_i - \mathbf{I}_i \in \mathbb{F}_2^{d_i \times d_i}$, where the matrix $\mathbf{1}_i$ is $d_i \times d_i$ block matrix of ones, and $\mathbf{I}_i$ is $d_i \times d_i$ identity matrix. The matrix $\mathbf{\Gamma} \in \mathbb{F}_2^{b \times b}$ is of the following block structure:
		\begin{equation}
		\begingroup\makeatletter\def\f@size{8}\check@mathfonts
		\mathbf{\Gamma} = \left( \begin{array}{cccc}
 		\mathbf{0}_{1,1} & \mathbf{\Gamma}_{1,2} & \dots & \mathbf{\Gamma}_{1,m}  \\
 		\mathbf{\Gamma}_{2,1} & \mathbf{0}_{2,2} & \dots & \mathbf{\Gamma}_{2,m}   \\
		\vdots & \vdots  & \hfill &\vdots   \\
		\mathbf{\Gamma}_{m,1} & \mathbf{\Gamma}_{m,2}  & \dots & \mathbf{0}_{m,m} \\
		\end{array} \right),
		\endgroup
		\label{}
		\end{equation}
where $\mathbf{0}_{i,i}$ is a block matrix $d_i \times d_i$ of zeros, and  $\mathbf{\Gamma}_{i,j} \in \mathbb{F}_2^{d_i \times d_j}$ with the $(i,j)$-th entry $\mathbf{\Gamma}_{i,j}(i,j)=$ $1$ if both $b_i^q$ and $b_j^q$ are incident to $x_q$ and $0$ otherwise. Note that holds $\mathbf{\Gamma}_{j,i} = \mathbf{\Gamma}_{i,j}^\mathrm{T}$.      

The vector of means $\mathbf {r}_{\mathrm{s}} \in \mathbb{R}^b$ can be decomposed as $\mathbf {r}_{\mathrm{s}}^{(\tau)} = [\mathbf {r}_{\mathrm{s},1}^{(\tau)}, \dots, \mathbf {r}_{\mathrm{s},m}^{(\tau)}]^{\mathrm{T}}$, where $\mathbf {r}_{\mathrm{s},i} \in \mathbb{R}^{d_i}$ is equal to $\mathbf {r}_{\mathrm{s},i}^{(\tau)} = [ r_{f_{i} \to \Delta x_k}^{(\tau)}, \dots, r_{f_{i} \to \Delta x_K}^{(\tau)}]$. Further, the vector $\mathbf{r}_{\mathrm{a}} = \big[\mathbf{r}_{\mathrm{a,1}}, \dots, \mathbf{r}_{\mathrm{a},m} \big]^{\mathrm{T}} \in \mathbb{R}^{b}$ contains means of indirect factor nodes, where $\mathbf{r}_{\mathrm{a},i} = [r_i, \dots, r_i] \in \mathbb{R}^{d_i}$. The diagonal matrix $\hat{\bm \Sigma}_{\mathrm{s}} \in \mathbb{R}^{b \times b}$ is obtained as $\hat{\bm \Sigma}_{\mathrm{s}} =$ $\lim_{\tau \to \infty} {\bm \Sigma}_{\mathrm{s}}^{(\tau)}$. The vector $\mathbf{r}_{\mathrm{b}} = \big[\mathbf{r}_{\mathrm{b},1}, \dots, \mathbf{r}_{\mathrm{b},m} \big]\in \mathbb{R}^{b}$ contains means from direct and virtual factor nodes to a variable node, where the $i$-th element $\mathbf{r}_{\mathrm{b},i} = \big[r_{\Delta x_k}, \dots, {r}_{\Delta x_K} \big] \in \mathbb{R}^{d_i}$. 

\textbf{Theorem 3 Proof:}
To prove theorem it is sufficient to show that \eqref{rand_5} converges to the fixed point defined in \eqref{fixed}. We can write:
		\begin{equation}
        \begin{aligned}
        \mathbf {r_r}^{(\tau-1)} &= 
        \mathbf {R}\mathbf{\widetilde r}
        - \mathbf {R} \bm \Omega
        \mathbf {r}_{\mathrm{s}}^{(\tau-2)}.
        \label{rand_6}
        \end{aligned}
		\end{equation}
Substituting \eqref{rand_1}, \eqref{rand_2} and \eqref{rand_6} in \eqref{rand_3}, and using that fixed point equals $\hat{\mathbf {r}}_{\mathrm{d}}=\lim_{\tau \to \infty} \mathbf {r}_{\mathrm{d}}^{(\tau)}$:		
		\begin{multline}
        \hat{\mathbf {r}}_{\mathrm{d}} = 
        \big( \mathbf {I} + \mathbf {Q} \bm \Omega + \alpha_2\mathbf {R} 
        \bm \Omega + \alpha_1\mathbf {R} \bm \Omega \big)^{-1} \\
        \cdot \big(\mathbf {Q}+  \alpha_2 \mathbf {R} 
        + \alpha_1 \mathbf {R}\big)
        \mathbf{\widetilde r}.     
        \label{rand_8}
		\end{multline}
From definitions of $\mathbf {Q}$, $\mathbf {R}$ and $\alpha_2$, we have $\mathbf {Q} \bm \Omega + \alpha_2\mathbf {R} \bm \Omega + \alpha_1\mathbf {R} \bm \Omega = \bm \Omega$ and $\mathbf {Q}+  \alpha_2 \mathbf {R} + \alpha_1 \mathbf {R} = \mathbf {I}$, thus \eqref{rand_8} becomes:
		\begin{equation}
        \begin{aligned}
        \hat{\mathbf {r}}_{\mathrm{d}} = 
        \big( \mathbf {I} + \bm \Omega \big)^{-1}\mathbf{\widetilde r}.  
        \end{aligned}
		\end{equation}
This concludes the proof.

\bibliographystyle{IEEEtran}
\bibliography{cite}

\end{document}